\documentclass[aps,prb,showpacs,twocolumn,superscriptaddress,nobibnotes]{revtex4-1}
\usepackage[utf8]{inputenc}
\usepackage{amsmath}
\usepackage{graphicx}

\usepackage{color}
\usepackage[dvipsnames, table]{xcolor}
\usepackage{algpseudocode,algorithm,algorithmicx}

\graphicspath{{images/}}

\newcommand{\bra}[1]{\langle\,#1\,|} 
\newcommand{\ket}[1]{|#1\rangle}
 
\newcommand{\braket}[2]{\langle\,#1\, | \, #2\,\rangle} 
\newcommand*\Let[2]{\State #1 $\gets$ #2}

\newcommand{\beginsupplement}{%
        \setcounter{table}{0}
        \renewcommand{\thetable}{S\arabic{table}}%
        \setcounter{figure}{0}
        \renewcommand{\thefigure}{S\arabic{figure}}%
     }
     
\begin{document}
\title{Variational optimization in the AI era: \\ Computational Graph States and Supervised Wave-function Optimization}

\author{Dmitrii Kochkov}
\affiliation{Institute for Condensed Matter Theory and Department of Physics, University of Illinois at Urbana-Champaign, IL 61801, USA} 

\author{Bryan K.  Clark}
\affiliation{Institute for Condensed Matter Theory and Department of Physics, University of Illinois at Urbana-Champaign, IL 61801, USA}

\begin{abstract}
Representing a target quantum state by a compact, efficient variational wave-function is an important approach to the quantum many-body problem. In this approach, the main challenges include the design of a suitable variational ansatz and optimization of its parameters.  In this work, we address both of these challenges. First, we define the variational class of Computational Graph States (CGS) which gives a uniform framework for describing all computable variational ansatz. Secondly, we develop a novel optimization scheme, supervised wave-function optimization (SWO), which systematically improves the optimized wave-function by drawing on ideas from supervised learning. While SWO can be used independently of CGS, utilizing them together provides a flexible framework for the rapid design, prototyping and optimization of variational wave-functions. We demonstrate CGS and SWO by optimizing for the ground state wave-function of 1D and 2D Heisenberg models on nine different variational architectures including architectures not previously used to represent quantum many-body wave-functions and find they are energetically competitive to other approaches. One interesting application of this architectural exploration is that we show that fully convolution neural network wave-functions can be optimized for one system size and, using identical parameters, produce accurate energies for a range of system sizes. We expect these methods to increase the rate of discovery of novel variational ansatz and bring further insights to the quantum many body problem.

\end{abstract}
\maketitle

\section{Introduction}
In the variational approach to the quantum many-body problem, one tries to find a compact efficient variational wave-function which is a good representation of the ground state. To accomplish this, one typically first selects a class of parameterized wave-functions and then optimizes over these parameters to find the best wave-function within this class.  

Early classes of wave-functions included Jastrow wave-functions used to represent Helium-4 \cite{mcmillan1965ground}  and Slater-determinants for Fermion systems \cite{slater1951simplification}.  Since this time there have been a multitude of other classes developed; examples include matrix-product states \cite{schollwock2011density,vidal2003efficient,white1992density}; projected BDG wave-functions\cite{capriotti2002projected}; Huse-Elser states \cite{huse1988simple,changlani2009approximating,mezzacapo2009ground,marti2010complete}; and restricted Boltzmann machine wave-functions \cite{carleo2017solving,deng2017machine,kaubruegger2018chiral,torlai2018neural, deng2017quantum, glasser2018neural, choo2018symmetries}. In addition to atomic ansatz, there has been significant work in combining variational wave-functions in various ways:  the Slater-Jastrow form \cite{martin2016interacting}, the product of a Slater and Jastrow wavefunction, is the most common ansatz used in ab-initio quantum Monte Carlo simulations; Slater-MPS \cite{chou2012matrix} and Slater-RBM \cite{nomura2017restricted} have improved on these approaches;  the multi-determinant ansatz \cite{olsen1988determinant}, consisting of sums of determinants, has been heavily used in chemistry and has seen a recent resurgence in the context of selected CI \cite{holmes2016heat} and as a constraint for auxilliary field QMC \cite{chang2017multi}.  Multi-Slater Jastrow (i.e. products of multi-determinants times Jastrow) \cite{bouabcca2010multi,morales2012multideterminant} improves upon these ansatz.  Neural network backflow \cite{luo2018backflow} and iterative backflow \cite{taddei2015iterative} techniques are essentially repeated function composition of the backflow techniques orginally pioneered by Feynman \cite{feynman1956energy}. We see that there is a large combinatorial explosion of different classes of wave-function possibilities. 

\begin{figure}
  \includegraphics[width=7cm]{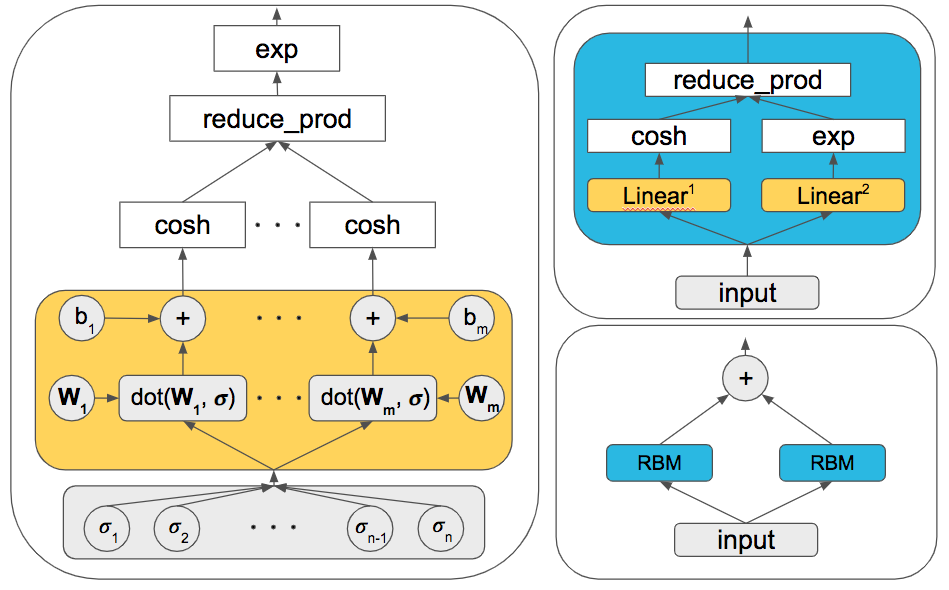}
  \caption{\textit{(left)} Computational graph representation of half of the restricted Boltzmann machine.  Linear transformation highlighted in yellow. \textit{(top right)} Computational graph representation of full RBM wave-function. Linear transformation (yellow) is now represented as an atomic operation.  \textit{(bottom right)} Representation of a computational graph shows composition of RBM to generate a multi-RBM. For all representations the wave-function amplitude is obtained by propagation of the corresponding input values through the graph, until the output node is reached.}
\end{figure}\label{fig:CGS_RBM}

 Ideally one would automatically choose both the class of wave-function and then the best parameters within this class. While this is not currently possible, an important first step in this direction is (1) a framework which places the plethora of variational ansatz on similar footing allowing quick prototyping, testing, and experimental design of variational wavefunctions and (2) effective algorithms to optimize them.  These two aspects will be the focus of this work. 
 
To accomplish this first task, we define a class of \textit{computational graph states (CGS)}  which encompasses the set of all computable wave-functions.   Wave-functions in this framework can be visualized as directed graphs connecting input nodes that represent configurations in the computational basis to a single output node that corresponds to the wave-function amplitude (see fig.~\ref{fig:CGS_RBM} for an example of this for RBM wave-functions). The structure of the graph includes variational parameters of the ansatz and transformations that produce the final amplitude. We can then think of various variational ansatz as simply different architectural choices for these computational graphs.
 
As this framework has been heavily used by the machine learning community, it most naturally encodes various machine-learning ansatz but also can be straightforwardly used for other models including matrix-product states and Slater-determinants. 
We highlight this freedom of CGS explicitly by constructing architectures for various different variational ansatz (see fig.~\ref{fig:CGS_RBM}, fig.~\ref{fig:architectures} and the supplement sec.~\ref{subsec:AdditionalArchitectures}).  
 
Once we have chosen an architecture, it is then important to optimize the parameters of that architecture.  There exists a number of methods for parameter optimization in the context of variational Monte Carlo. Here we introduce an additional technique, \textit{supervised wave-function optimization (SWO)} which combines ideas from machine learning and traditional wave-function optimization.  Unlike many traditional optimization schemes which walk downhill in energy, SWO minimizes the difference between the current parameterization and an explicit improved wave-function.   While SWO can be instantiated in various ways, one approach implements imaginary time evolution (IT-SWO) in a way that scales independently of the number of variational parameters.

\section{Computational graph states}

Computational graphs lie at the heart of the dataflow programming model \cite{abadi2016tensorflow, al2016theano, akidau2015dataflow}. In this paradigm arbitrary computation is represented as a directed graph composed of a set of nodes. Each node has zero or more incoming and outgoing edges that represent the dataflow and explicitly define the evaluation order. Nodes represent operations, which upon execution generate output based on the inputs and pass it along the outgoing edges. Examples of operations include insertion of runtime inputs (typically provided by the user), retrieval of stateful variables (injection of trainable weights) and mathematical transformation (Add(x,y)). The result of the computation is obtained by propagating data through the graph until the value of the requested node is obtained. The class of Computational Graph States (CGS) uses this paradigm transforming an input configuration to an output amplitude. 
The efficiency and representation power of a particular architecture is determined by the connections and nodes of the graph which can be arranged to represent arbitrary computation.

As an example, let us start by showing how we can represent a RBM wave-functions (RBM-WF) as a CGS. RBM-WF are typically represented as Ising models with bipartite interactions between visible spins (representing the configuration of your quantum system) and hidden spins. The amplitude in a RBM-WF for a given configuration is the probability of the Ising configuration at fixed visible spins integrated over all the configurations possible for the hidden spins. This is not a feed-forward architecture and so doesn't directly translate into a computational graph. Instead, we should ask what computational steps are taken to efficiently compute the amplitude. In this case, because of the bipartite nature, the amplitude of a given configuration is equal to 

\begin{equation}
\left[ \prod\limits_{i}\cosh\left(\sum\limits_{j=1..N}W_{ij}\sigma_{j} + b_{i} \right)\right] \exp\left(\sum\limits_{j} a_{j}\sigma_{j} \right)
\end{equation}

This algebraic formula can be readily converted into a sequence of operations. Expression in the square brackets breaks the process into 4 steps: evaluation of the \textit{inner product}, \textit{addition}, \textit{cosh}, \textit{product reduction}. These steps are shown in a computational graph form in fig~\ref{fig:CGS_RBM}(\textit{left}). In fig.~\ref{fig:CGS_RBM}(\textit{top right}) we show a full expression where we have combined all weighted terms into 2 linear transformations over the input vector.  Note that in the computational graph framework it is easy to efficiently add specialized operations, such as cosh, the determinant of a matrix, convolutions, etc.

We can expand the RBM-WF ansatz in various ways.  For example, one can add nonlinear transformation on inputs (see the FC-RBM in fig.~\ref{fig:architectures}).  In addition, when represented as a computational graph, it is easy to create a sum/product structure over wave-functions by combining different computational blocks through an addition or multiplication node.  For example, we can build a multi-RBM by summing two RBM structures (see fig.~\ref{fig:CGS_RBM}(\textit{bottom right})).

It is also often straightforward to explicitly put in constraints (i.e. translational invariance) in such graphs decreasing the number of parameters which have to be optimized and encoding physical knowledge.  For example, in the context of an RBM, this could be accomplished by replacing the linear transformation blocks with convolution blocks.

Computational graph architectures come with a handful of favorable properties for quick prototyping and efficient implementations.  As a central tool in machine learning research, multiple frameworks~\cite{abadi2016tensorflow,al2016theano,jia2014caffe,seide2016cntk} implement automatic differentiation for computational graphs, which generates operations that evaluate derivatives with respect to requested nodes; automatic differentiation has previously been used in VMC in the context of computing forces \cite{sorella2010algorithmic}.  Automatic differentiation eliminates the necessity to hand-code gradients with respect to all variational parameters, which is often an error-prone and time-consuming procedure that greatly limits the scope of architectures typically considered in VMC. Computational graph structure provides a flexible way to inject domain knowledge and experiment with novel computational units and architectures, which has been a major source of advances in the field of machine learning and artificial intelligence \cite{lecun1990handwritten,hochreiter1997long, he2016deep,szegedy2015going,defferrard2016convolutional,krizhevsky2012imagenet,graves2016hybrid,van2016wavenet}. It has been found to be especially important for simulation related applications \cite{bar2018data, tompson2016accelerating, xie2018tempogan, de2017deep, rasp2018deep}.

In addition to theoretical benefits, the dataflow programming paradigm is well suited for distributed execution and optimization. Modern tools provide heuristics for evaluation scheduling and support of specialized hardware accelerators (GPUs, TPUs). Having computational graphs as a central part of this work, we have implemented the entire process, including sampling, evaluation and training in TensorFlow (see sec.~\ref{sec:TensorFlow} for details).

In addition to efficiently representing wave-function architectures, it is important to be able to efficiently optimize variational parameters. Similar to machine learning applications, some CGS instances reside in the regime where the number of parameters is much larger than that admissible by second order methods. In the next section, we present a novel wave-function optimization algorithm inspired by supervised learning.

\section{Supervised Wavefunction Optimization (SWO)}
Here we present a new optimization method, \textit{Supervised Wave-Function Optimization} (SWO).  Like other optimization schemes, SWO tries to find the best parameters $W$  in a space of wave-functions $\ket{\Psi_W}$. In this work, we are thinking of the best parameters as those which are closest to the ground state wave-function of a given Hamiltonian $\hat{H}=\sum_k \hat{h}_k$ but SWO naturally generalizes to a variety of other cases such as time-evolution and eigenstates; related ideas using wave-function matching for simulating quantum circuits have been used in a recent independent work \cite{jonsson2018neural}. SWO works over a number of epochs where at each epoch, SWO updates $W$ to minimize the  difference between $\ket{\psi_W}$ and a (epoch-specific) target wave-function $\ket{\psi_{T}}$. This minimization is cast as a supervised machine learning task which `learns' the target wave-function from a set of configurations $\{c\}$ chosen with probability $p(c)$ which are `labelled' with their target wave-function amplitudes $\braket{c}{\Psi_T}$.
 
This minimization forms the inner loop of SWO, which we perform in multiple steps. At every step the parameters are tuned by gradient descent on the loss function $L = \mathcal{L}_{2}(\epsilon_{rr})$ where $\epsilon_{rr} \equiv (\psi_{W}-\psi_{T})$. To leading order in $\epsilon_{rr}$, this is equivalent to optimizing the fidelity $f_{err}  = 1 - \braket{\psi_{T}}{\psi_{W}} /\sqrt{|\psi_{T}|^{2}|\psi_{W}|^{2}}$ (see sec.~\ref{sec:SuppFidelity} for more details and an alternative objective function). For each step $i$ (starting from 0), we generate a set of configurations with probability $p(c) \propto |\Psi_{Wi}(c)|^2$ (where $\Psi_{W0} \equiv \Psi_W$). Then, we evaluate $\braket{c}{\Psi_T}$ and $\braket{c}{\Psi_{Wi}}$.  A stochastic estimation of $L$ is then given by

\begin{figure}[H]
  \includegraphics[width=8cm]{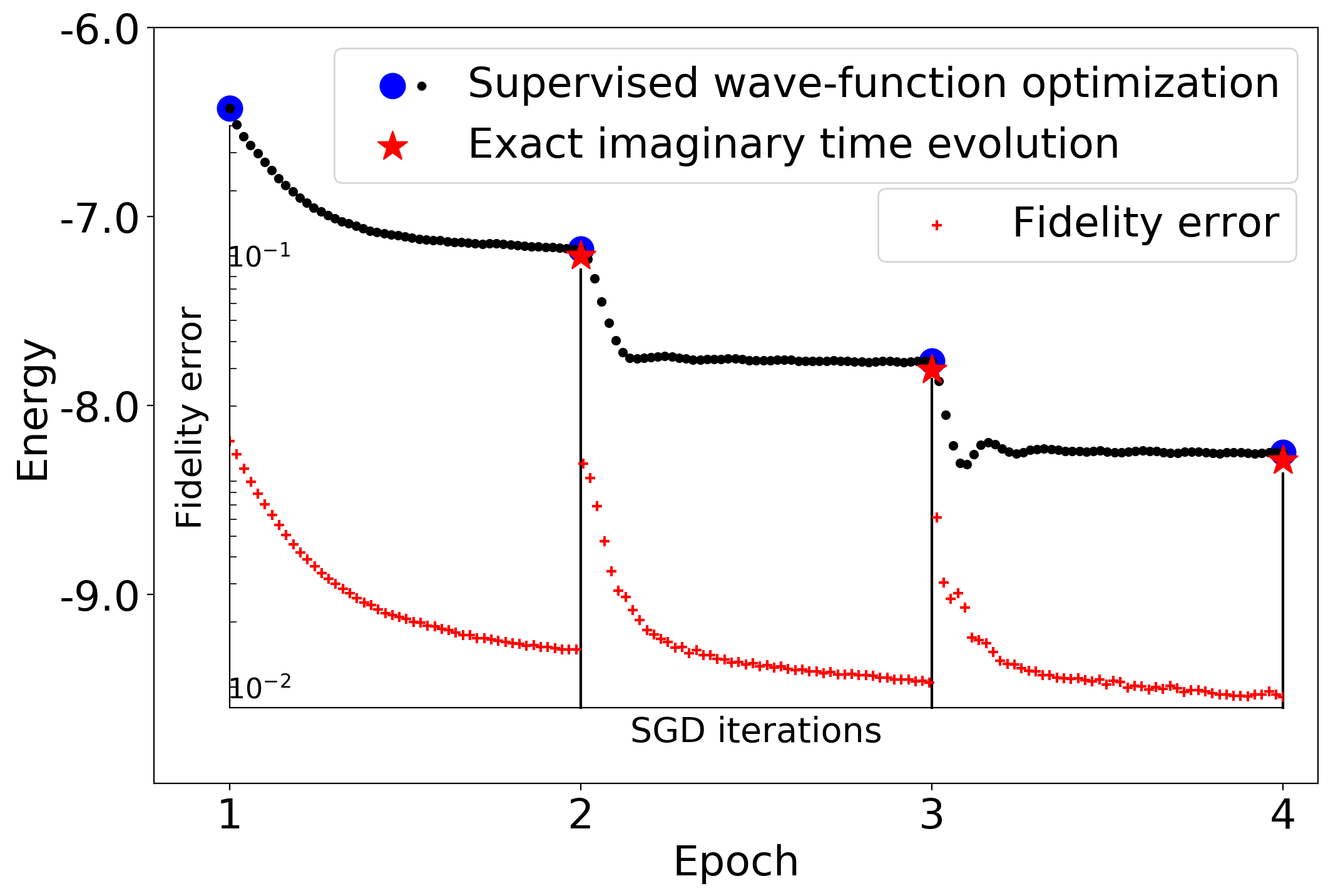}
  \caption{The process of SWO optimization demonstrated on a 24 site 1d Heisenberg model using a FCNN (2 hidden layers, 80 neurons). \textit{Main panel:}  For each epoch $i$ the energy is shown for the wave-function exactly time-evolved from the SWO state at the previous epoch (red stars) and the SWO approximation of time-evolution from the same wave-function (blue dots).  Each SWO epoch consists of multiple steps whose respective energy is shown in black.  Note this energy need not be monotonic.  \textit{Embedded plots:} Fidelity error (red dots) between the target wave-function generated from exact imaginary time evolution and the state created by SWO.}
  \label{fig:time_evolution}
\end{figure}

\begin{figure*}
  \includegraphics[width=14cm]{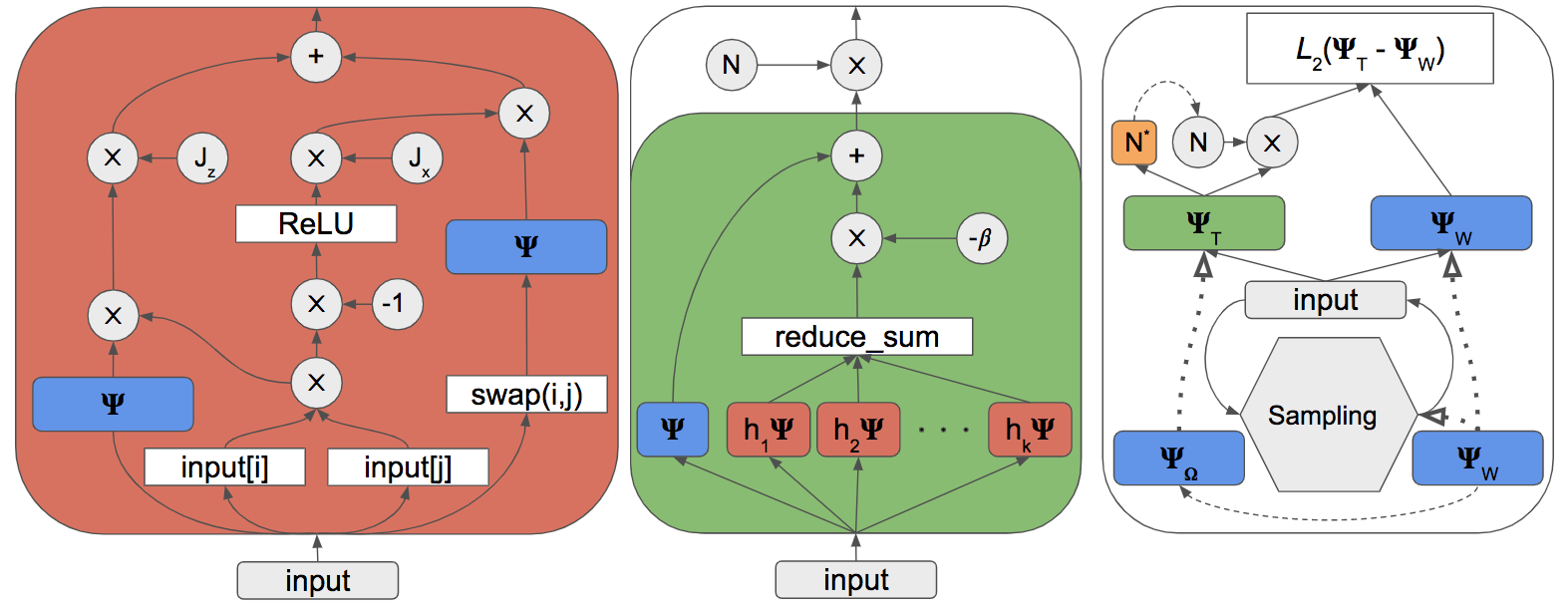}
  \caption{
    Building blocks of SWO optimization algorithm. (Left panel) Computational graph representation of $(J_{z} \hat{S}_{i}^{z} \hat{S}_{j}^{z} + J_{x} \hat{S}_{i}^{x,y} \hat{S}_{j}^{x,y}) \ket{\psi_{W}}$. Square bracket operation represents slicing and swap operation along the spin index dimension. Blue $\Psi$ block represents computational graph of the wave-function. (Central panel) Computational graph representation of $(1-\beta \hat{H})\ket{\psi_{W}}$. Red blocks $h_{i} \Psi$ represent wave-function after application of individual terms, similar to Left panel. (Right panel) Computational graph showing the high level operation of the SWO algorithm. The top part shows the process of calculation of $\mathcal{L_{2}}$ norm, which is used for estimation of the gradients with respect to variational parameters. The bottom part generates samples on which gradients are evaluated. Dashed lines show operations executed once per epoch and dotted line indicates the weight sharing. Blocks of similar structure are color-coded.
  }
  \label{fig:target_state_construction}
\end{figure*}

\begin{equation}
L \approx \sum_c \frac{(\braket{c}{\Psi_T} - \braket{c}{\Psi_{Wi}})^{2}}{p(c)}.
\end{equation}

Using this stochastic estimate (and derivatives thereof), we then take a step in parameter space to decrease $L$ returning the improved wave-function $\Psi_{W(i+1)}$.  In practice we use an adaptive SGD technique, such as Adam, where momenta are kept between epochs. We keep taking steps until $L$ is sufficiently small thereby completing one epoch of the algorithm.  A new improved target wave-function is then chosen for the next epoch (see fig.~\ref{fig:time_evolution}).

There are various approaches to get a better target wave-function $\ket{\psi_T}$. The only requirement is the ability to evaluate the (unnormalized) amplitude $\langle c | \psi_T\rangle$. In some sense, an ideal situation is to have a target ground state to match; while this is obviously impractical, one can emulate something close to this by matching against an easier-to-train architecture which is close to the ground state (matching-SWO). We show example of doing such in sec.~\ref{subsec:match}. A more systematic way of generating improved wave-functions is through the application of a propagator which brings you closer to the ground state. Examples include $\exp[-\tau \hat{H}]$ (potentially trotterized), $(1-\tau \hat{H})$, Lanczos steps \cite{sorella2001generalized}, etc. In this work, we primarily work with $\ket{\psi_{W}} -\beta  \hat{H} \ket{\psi_{W}}\equiv \ket{\psi_T}$,  which is approximately $e^{-\beta \hat{H}}\ket{\psi_{W}}$ at small $\beta$ so as to  more directly contrast the algorithmic steps in SWO with the algorithmic steps of other methods for imaginary time evolution \cite{sorella2001generalized,vidal2004efficient}. We label this specific instantiation of SWO the imaginary time implementation (IT-SWO).

Due to the non-unitary nature of imaginary time evolution, the norm of $\ket{\psi_{T}}$ is not preserved. To avoid  learning an irrelevant new normalization, we introduce a normalization factor \textit{N} that is updated during training. The value of normalization is determined by
\begin{align*}
    \braket{\psi_{T}}{\psi_{T}} = &N^{2}\bra{\psi_{W}}(1 - 2\beta \hat{H} + \beta^{2} \hat{H}^{2}\ket{\psi_W} = \braket{\psi_{W}}{\psi_{W}} \\
    &N = (1 - 2 \beta \langle E \rangle + \beta^{2} \langle E^{2} \rangle)^{-0.5}
\end{align*}
We compute $N$ for a given epoch by an exponential moving average (EMA) of the steps from previous epochs.

\begin{algorithm}[H]
    \caption{Supervised Wave-function Optimization \\ (Imaginary Time Implementation)\label{alg:optimization}}
    \begin{algorithmic}[1]
        \Require{$\beta$, $num\_epochs$, $num\_sgd\_steps$}
        \State{Initialize $c$}
        \State{Initialize $\psi_{W_\alpha}$}
        \State{Initialize $\psi_{T}=\psi_{W_{\Omega}} - \beta \hat{H} \psi_{W_{\Omega}}$}
        \State{Initialize $N=N^{*}=(1 - 2\beta \langle E \rangle + \beta^{2} \langle E^{2} \rangle)^{-0.5}$}
        \For{$i \gets 1 \textrm{ to } num\_epochs$}
        \Let{$W_{\Omega}$}{$W_{\alpha}$}
        \Let{$N$}{$N^{*}$}
        \For{$j \gets 1 \textrm{ to } num\_sgd\_steps$}
        \State{Minimize$[\mathcal{L}_{2}(N\psi_{T}(c) - \psi_{W_{\alpha}}(c))]$}
        \State{$N^{*} = EMA((1 - 2\beta \langle E \rangle + \beta^{2} \langle E^{2} \rangle)^{-0.5})$}
        \Let{$c$}{$MCMC(c, \psi_{W_{\alpha}})$}
        \EndFor
        \EndFor
    \end{algorithmic}
\end{algorithm}

It is worth noting that the variance of the derivatives of the objective function go to zero as $\Psi_{W\alpha}$ approaches $\Psi_T$ (even if $\Psi_T$ is not the true ground state). This is commonly known in the literature as a zero-variance estimator for the derivatives.

There are two parameters to tune in IT-SWO:  the value of $\beta$ and the number of steps per epoch.  In principle, the most conservative approach is to use a large number of steps per epoch ensuring a high fidelity approximation to the imaginary time path; we find this does minimize the number of total epochs required as expected (see fig.~\ref{fig:steps_per_epoch}(\textit{left})).  Unfortunately, this makes each epoch slower and since most of the fidelity improvement happens in the first few steps (see fig.~\ref{fig:time_evolution}) often a more aggressive approach is useful in practice  to minimize the wall-clock time (see fig.~\ref{fig:steps_per_epoch}(\textit{right})).   This is further discussed when comparing IT-SWO with other algorithmic approaches.

\subsection{SWO implemented as a computational graph}
The question now becomes how one explicitly generates training samples, target wave-function amplitudes and gradients with respect to variational parameters needed for optimization. For our implementation, we implement all of the components of SWO as computational graph as seen in fig.~\ref{fig:target_state_construction}(c), which later is repeatedly evaluated in the order described in algorithm \ref{alg:optimization}. In this framework evaluation of gradients is automated by virtue of symbolic differentiation, enabling us to work with much more complicated ansatzs and employ various training heuristics such as Nesterov's accelerated gradient descent \cite{nesterov1983method}, momentum\cite{sutskever2013importance} , Adam \cite{kingma2014adam},  etc.

To construct the computational graph representing the target state $\ket{\psi_{T}} = \ket{\psi_{W}} - \beta \sum_{k}\hat{h}_{k} \ket{\psi_{W}}$ we use the composite property of computational graphs. It suffice to stitch together modules representing individual terms $\left\{\ket{\psi_{W}}, \hat{h_{k}} \ket{\psi_{W}}, .. \right\}$ with appropriate coefficients. Implementation of a single term $\hat{h}_{k} \ket{\psi_{W}}$ is broken into primitive components by insertion of an identity operator $\sum_{\sigma^{'}} \ket{\sigma^{'}} \bra{\sigma^{'}}$ into $\bra{\sigma} \hat{h_{i}} \ket{\psi_{W}} = \sum_{\sigma^{'}} \bra{\sigma} \hat{h_{i}} \ket{\sigma^{'}} \braket{\sigma^{'}}{\psi_{W}}$. For local Hamiltonian individual terms $\bra{\sigma} \hat{h_{k}} \ket{\sigma^{'}}$ involve only a few non-zero matrix elements and can be efficiently represented as computational graphs. In fig.~\ref{fig:target_state_construction}(a) we give an explicit example for a generalized Heisenberg bond operator $\hat{h}_{k} = J_{z} \hat{S}_{i}^{z} \hat{S}_{j}^{z} + J_{\perp}( \hat{S}_{i}^{x} \hat{S}_{j}^{x} + \hat{S}_{i}^{y} \hat{S}_{j}^{y})$. 
All terms combined into a final wave-function are shown in fig.~\ref{fig:target_state_construction}(b).  We then generate computational graphs for the entire SWO procedure. Sampling is done by building a computational graph for  Markov chain Monte Carlo and the objective function is a trivial computational graph which takes the averaged squared difference between target and current wave-functions (see fig. \ref{fig:target_state_construction}(c)).

\subsection{Computational complexity} 

Here we consider the computational complexity of using SWO as an optimization technique. Per configuration, let us define the cost of evaluating a wave-function as $O(C)$, the cost of evaluating the target wave-function as $O(C')$ and the cost of evaluating the local energy as $O(E)$.  Beyond a typical non-optimizing variational Monte Carlo simulation, the additional cost of SWO is determining the gradient of the $L_2$ norm for each configuration (selected once per sweep) with respect to all the parameters.  Per configuration, this costs $O(C'+C)$ where the $O(C')$ cost comes from evaluating the target wave-function and,  by the virtue of automatic differentiation, $O(C)$ is the cost of computing the derivatives of the $L_2$ norm with respect to all the parameters. 
In the particular instance of imaginary time evolution, evaluation of $\braket{C}{\psi_{T}}$ costs the same $O(E)$ as calculating the local energy  giving a total cost per configuration (and therefore sweep) of $O(C+E)$.  In addition to the cost of computing the derivatives of the $L_2$ norm, there is the standard variational Monte Carlo cost of sampling and computing the local energy which goes naively as $O(CN+E)$ per sweep.  For some wave-functions, such as projected BDG, update formulas bring this VMC sampling cost down to $O(C+E)$.  Per sweep, it is surprising then that SWO scales just as the cost of computing the energy of a fixed wave-function with essentially no dependence of the number of parameters.  Above we have determined the per sample cost.  The number of sweeps required for SWO will be system-dependent and may depend on the number of parameters in the wave-function. To get an estimate, for the simulations reported in this work we are primarily using 200 sweeps per mini-batch and 400 mini-batches per epoch. The simulations reported here took roughly a minute per epoch.

\subsection{Comparison to other methods}
\begin{figure}
  \includegraphics[width=7cm]{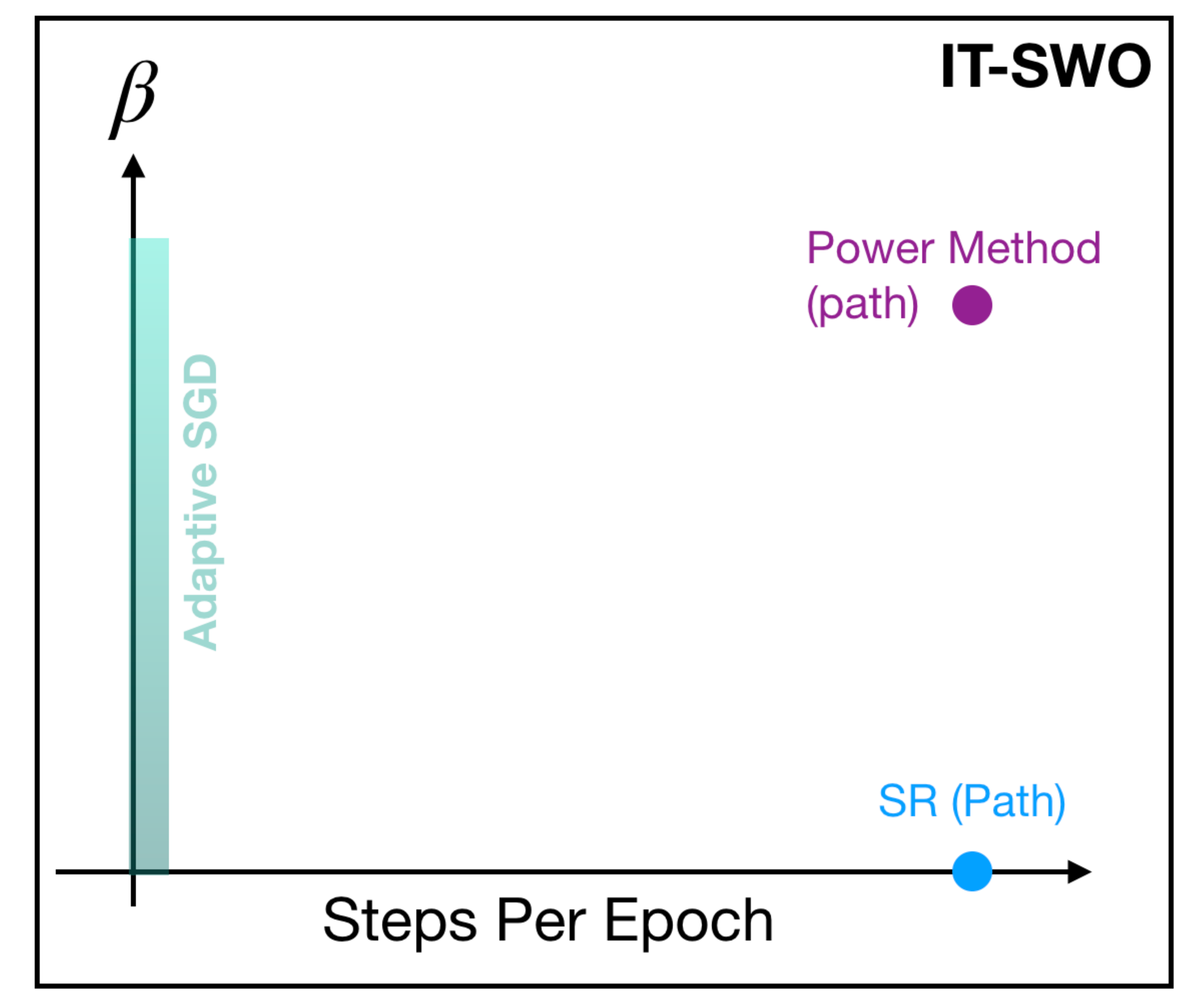}
  \caption{Illustration showing how tuning the parameters of IT-SWO allows it to interpolate between adaptive SGD on the energy and an algorithm which follows the SR or power method path.  At large $\beta$ and a single step per epoch there is a $\beta^2$ deviation from adaptive SGD coming from the standard approach to the normalization. 
  }
  \label{fig:IT-SWOComparison}
\end{figure}

When run at small $\beta$, by tuning the number of steps per epoch, IT-SWO interpolates between two promising optimization techniques: adaptive stochastic gradient descent (SGD) and the stochastic reconfiguration path (sans the typical SR complications) (see fig.~\ref{fig:IT-SWOComparison}).  In addition to these interpolated limits of IT-SWO we also contrast it against imaginary time evolution with matrix-product states and the linear method.

Stochastic gradient descent (SGD) takes steps in parameter space in the stochastic gradient of the expectation of the energy.  It is efficient per step as the gradients can be computed quickly but converges slowly making it a poor choice for optimization of variational wave-functions although recent work \cite{schwarz2017projector,sabzevari2018faster} has shown that using acceleration methods improves on this situation. Stochastic reconfiguration (SR) \cite{sorella2001generalized} changes the parameters to most closely match $(1-\tau H)|\Psi\rangle$ following an approximate imaginary-time evolution that is believed to be a favorable optimization trajectory. To accomplish this, for a particular vector $\vec{v}$, it needs to compute (implicitly or explicitly) $S^{-1} \vec{v}$ where $S$ is the overlap matrix of parameters, $S_{ij} = \langle \partial \Psi/\partial i | \partial \Psi /\partial j\rangle$. Working with $S$ causes a number of problems including naively inefficient scaling; $S$ can be singular (especially when there are many parameters); a non-linear bias introduced by taking an inverse; and an undersampling problem when $|\Psi(c)\rangle $ and $|\partial \Psi/\partial \alpha(c)\rangle$ have large weight on significantly different configurations. While significant work has gone into attenuating many of these difficulties (i.e. $S$ can be regularized by adding a small value to the diagonal; the naive scaling can be improved by using iterative techniques and storing all the outer products), many of these problems do not arise in IT-SWO.

At small $\beta$, IT-SWO and SR both reach the best representation of $(1-\beta H)|\Psi_W\rangle$ for the variational manifold but in different ways.  SR solves directly for this state while IT-SWO finds it through multiple discrete SGD steps; IT-SWO therefore avoids (implict or explicit) evaluation of the overlap matrix solving most of the technical difficulties with SR. Under-sampling is resolved in SWO because samples for matching are selected from the immediately proceeding iteratively updated wave-function. This self-consistently ensures that the samples being considered are taken from (our best possible parameterized approximation) of the target wave-function. One could even further reduce this effect by sampling directly from the target wave-function but, at least in the IT version of SWO, this is significantly more expensive.

At larger $\beta$, IT-SWO and SR fundamentally diverge.  SR, with a large $\beta$, simply continues in the same direction as the infinitesimal imaginary time evolution.  IT-SWO instead finds the best finite-$\beta$ representation of $(1-\beta H)|\Psi_W\rangle$.  This means IT-SWO can be used at a $\beta$ which, instead of applying imaginary time evolution, is better described as applying the power method for finding extremal eigenvectors,  where the component of the ground state is increased iteratively through the application of $(H-\lambda)$ where $\lambda$ is chosen to make $|E_\textrm{min}|>|E_\textrm{max}|$.  In fig.~\ref{fig:varying_beta} we show the improved convergence per epoch of IT-SWO as we increase $\beta$. Since the steps per epoch required increases with $\beta$,  it is an open question how to choose the optimal $\beta$. The power method perspective suggests additional extensions based on accelerated iterative methods \cite{xu2018accelerated} although we do not yet see improvements applying this.

In the limit where IT-SWO takes only a single step per epoch, we reproduce adaptive (i.e. ADAM) SGD on the energy.  While more steps per epoch seems to consistently decrease the number of epochs needed, the wall-clock time for convergence is more complicated and we see cases where using adaptive SGD (i.e IT-SWO at 1 step per epoch) has the smallest wall-clock time.
Convergence improvements for variational wave-function optimization using adaptive SGD is consistent with what is seen in other recent studies \cite{schwarz2017projector,sabzevari2018faster}. It is an interesting open question determining, for each wave-function and Hamiltonian, what are the most efficient parameters for IT-SWO.

Another popular technique in VMC optimization, the linear method \cite{toulouse2007optimization,umrigar2007alleviation,neuscamman2012optimizing}, is harder to compare directly;  it too requires the computation of $S$ as well as the Hamiltonian in the tangent subspace. This comes along with a number of known issues including non-linear bias.  On the other hand it requires significantly fewer changes of the parameter, albeit many more sweeps per parameter, than techniques such as IT-SWO or SR and trade-offs between these approaches is left as an important open question particularly for highly non-linear architectures where a linearization of the parameters is a poor approximation.

A final approach worth comparing is imaginary time evolution in matrix-product states. Although there are different variants of this approach, one technique mirrors IT-SWO by generating the imaginary time-evolved wave-function and then matching the MPS in the variational manifold against it. This is analogous in many ways to IT-SWO but restricted to the MPS variational class.

While we have compared IT-SWO to various other optimization schemes in this section, it should be remarked that the SWO class of techniques spans significantly beyond imaginary time evolution. For example, one may be able to find states closer to the ground state by using a quicker to optimize but worse ansatz as an initial target wave-function or using Lanczos steps. 

\begin{figure*} \label{fig:swo_optimization}
  \includegraphics[width=7cm]{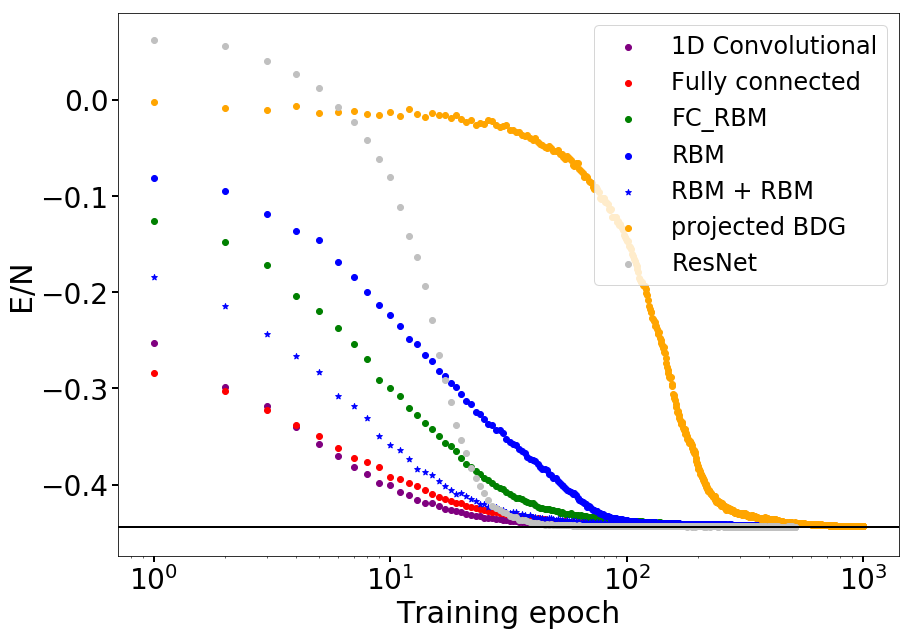}
  \includegraphics[width=7cm]{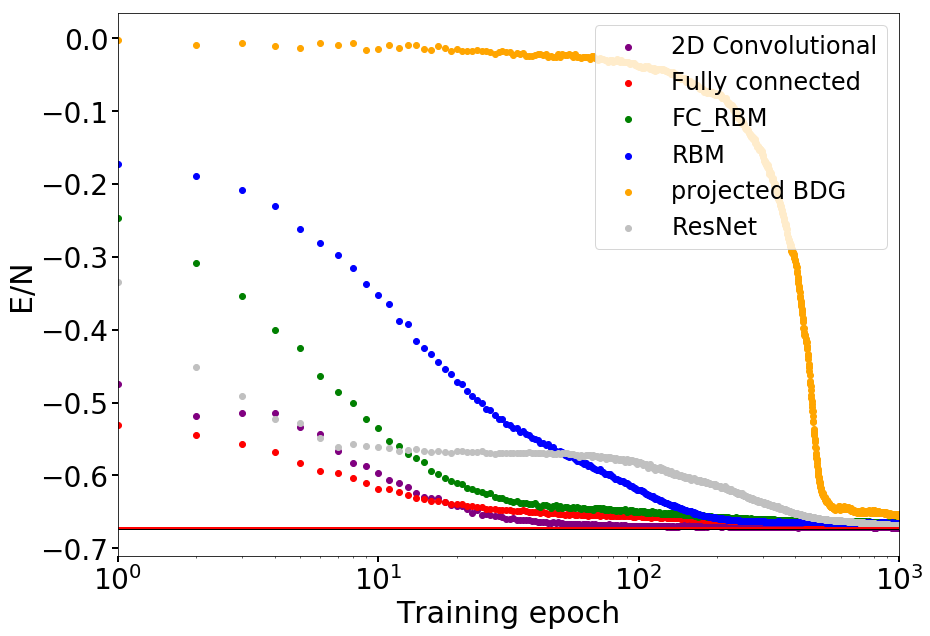}
  \caption{
    Energy per site as a function of training epoch (i.e. time evolution steps) for different variational ansatz during SWO optimization on (\textit{left}) 40 sites 1D Heisenberg model with periodic boundary conditions and (\textit{right}) 64 site square lattice Heisenberg model with periodic boundary conditions.  Different architectures are optimized using the same hyper-parameters except for ResNet which has a 10 times lower learning rate.
  }
\end{figure*}

\section{Examples}
\begin{table}[]
\begin{tabular}{|c|c|c|c|}
\hline
\textbf{WF} & \textbf{Hidden}  & \textbf{Hidden} & \textbf{Params}\\
   & \textbf{Layers}  & \textbf{Units}  &   1d - 2d      \\
\hline
FCNN           & 2   & 80                      & 9800 - 11744\\
\hline
RBM          & 0   & 80                        & 3320 - 5264\\
\hline
FC-RBM       & 2   & 80                        & 16280 - 18224 \\
\hline
Conv1D       & 5   & 16 filters (size 5)     & 5280 - \\
\hline
Conv2D       & 5   & 16 filters (5 by 5)       & - 26080\\
\hline 
ResNet       & 1 + 2(2)  & 16 filters (size 5)      & 5280 - 25760 \\
\hline
Multi-RBM   &  2   & 16 filters (5 by 5)    & 6640 - 10528 \\
\hline
Multi-FCNN       &  2   &    80     & 19600 - 23488 \\
\hline
P-BDG       &    &                &  1600 - 4096\\
\hline
MPS       &     &          &  4800 - \\
\hline

\end{tabular}
\caption{Parameters of architectures used in this work.}
\label{parameters_table}
\end{table}

\begin{figure}
  \includegraphics[width=7cm]{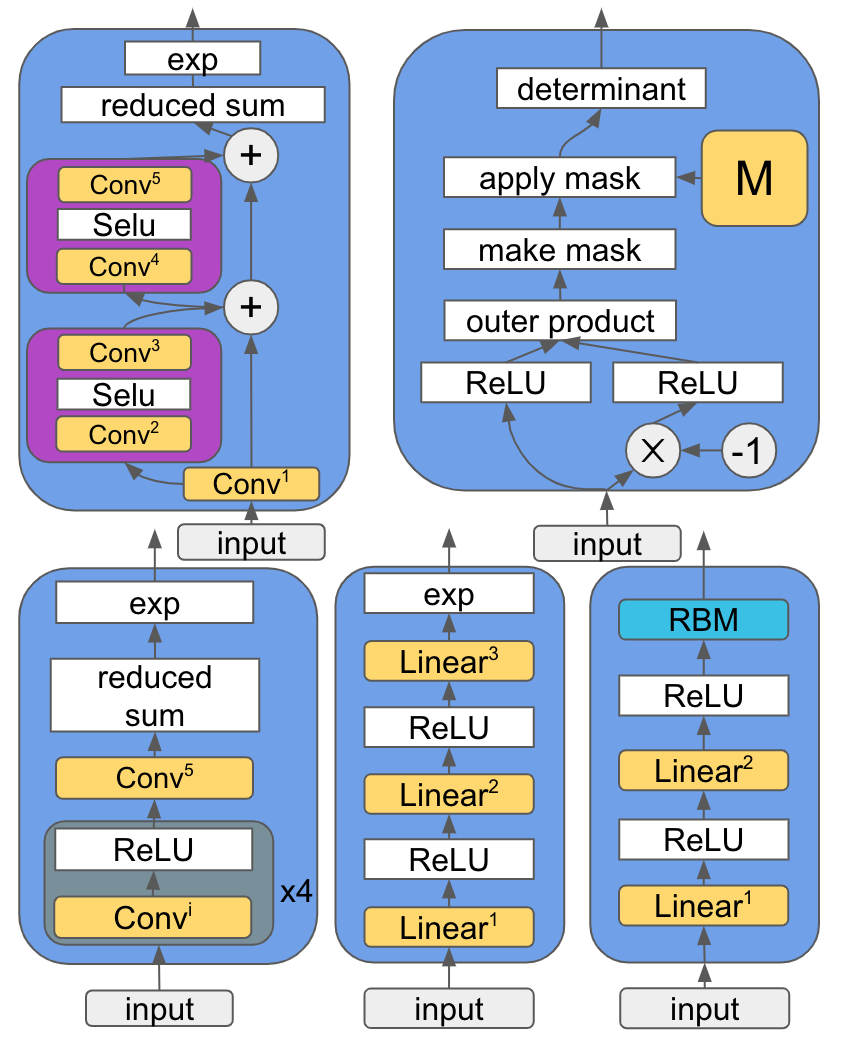}
  \label{fig:architectures}
  \caption{CGS for additional architectures (not including the RBM and multi-RBM in fig.~\ref{fig:CGS_RBM}) considered in the main text of this work for \textit{(top left)} ResNet, \textit{(top right)} projected BDG, \textit{(bottom left)} convolution networks, \textit{(bottom center}) FFNN, and \textit{(bottom right)} FC-RBM.  We also consider multi-FCNN and matrix-product states in the supplement (see~\ref{subsec:AdditionalArchitectures}). The graphs we use sometimes differ from those shown in technical ways to avoid underflow (typically by taking logarithms and then exponentiating later)
  }
\end{figure}

In this section, we construct and optimize via SWO a variety of CGS (see \ref{fig:architectures}) using Heisenberg models on a 1D chain and a 2D square geometry (as is typical \cite{carleo2017solving} in such benchmarks, we explicitly remove the Marshall sign rule).  The ansatz architectures we consider in the body of this paper include a generalized  projected BDG wave-function (P-BDG), restricted Boltzmann machine (RBM), multi-RBM, convolutional neural networks (CNN), a fully connected neural network (FCNN), the fully-connected RBM (FC-RBM), and a residual network (ResNet). All of these architectures except for P-BDG ansatz are capable of representing arbitrary quantum states given enough variational parameters, but different architecture design provides different convergence and fixed-parameter representability properties. In sec.~\ref{subsec:AdditionalArchitectures} we also show examples for matrix-product states. While some of these networks have been previously used in variational studies, many of them have not previously been used for representing wave-functions.

To build computational graphs, we follow the modular design pattern, where elementary components are reused in multiple architectures. We extensively use a `linear' module to introduce variational parameters that represents a linear mapping from $N$ dimensional input to $M$ dimensional output (see the yellow box in fig.~\ref{fig:CGS_RBM}). The other common form of linear transformations extensively used in AI and Machine Learning fields is convolutions, which can be interpreted as a linear transformation with a particular weight sharing \cite{lecun1989backpropagation, krizhevsky2012imagenet}. In addition to linear transformation we apply nonlinear transformation such as Selu, ReLU and `exp', that act on inputs element-wise. Reduction methods are used to collect contributions with equal weights.

From combining these simple building blocks we are able to build up the various ansatz mentioned above (see fig.~\ref{fig:architectures} and fig.~\ref{fig:CGS_RBM}).  Here we point out some important aspects of these architectures.    In our FCNN we alternate linear and activation blocks using ReLU as our activation function for all but the final layer where we instead use `$\exp$'.  While ReLU are standard for FCNN and other variational fully connected neural network work has used sigmoids \cite{cai2018approximating}, the use of an exponential activation function at the top is atypical and was selected based on the observation that the majority of ground state wave-function amplitudes take near-zero values. Exponential activation function provides a simple mechanism for learning this distribution, compared to naive linear outputs where a cancellation in the final output is necessary for an exponentially large number of input configurations.  Our FC-RBM and multi-RBM exemplifies how it is straightforward to combine architectures. In the FC-RBM we insert a series of alternating linear and ReLU blocks between the input and the rest of the RBM making the RBM architecture deeper.  In the multi-RBM we sum two RBM wave-functions in the spirit of multi-determinants. In the  CNN model, we use weight sharing to add translational invariance to our ansatz. ResNet expand around the CNN model by including connections from previous layers; such networks are designed to iteratively learn the remaining error in the log of the wave-function. Both of these architectures feature an exponential activation at the end similarly to FCNN. 

We have included (a superset of) projected BDG wave-function to demonstrates flexibility of the computational graph framework. Its computational graph consists of the determinant of matrix $A$ which is generated by taking a matrix $M$ of $n \times n$ variational parameters which are sliced and reshaped into a square $n/2 \times n/2$ matrix $A \equiv M[\{u\},\{d\}]$ where $\{u\}$ ($\{d\}$) are the lists of indices which specify the positions of the $n/2$ spin-up (spin-down) electrons.  Because we are using an arbitrary matrix  $M$ which is not guaranteed to be symmetric this is, strictly speaking, a superset of projected BDG states.  The product structure of the wave-function, similar to exponentials, provide an easy way of selecting the important configurations in the Hilbert space.

\subsection{IT-SWO}
In our IT-SWO optimization we use the same hyperparameters (except for ResNet) for all simulations to show that one could do reasonably well over many architectures  without fine-tuning. We used learning rate of 0.001 (0.0001 for ResNet), $\beta=0.12$, mini-batches of size 200 and 400 mini-batches per epoch. The optimization speed can be considerably improved by tuning hyperparameters and IT-SWO execution regime (see sec.~\ref{sec:SuppOptimization} for details).  Because here we are doing imaginary time simulations, if the ansatz are sufficiently expansive they should all follow the imaginary-time trajectory. In practice, because of the projection back into the variational manifold the paths will be different. Note also the different architectures start with different `random` initial conditions.

Figure~\ref{fig:swo_optimization} shows numerical optimization of these various architectures (see Table~\ref{parameters_table} for details such as number of layers and neurons per layer) for the one and two-dimensional system.  We find that IT-SWO is able to successfully optimize all architectures reaching energies close to the ground state in both Hamiltonians.   While we don't want to put too much emphasis on which architectures is best given that the sizes aren't directly comparable and the hyperparameters aren't individually optimized, we can still make some general observations about convergence to the ground state. In our benchmark we see that the projected BDG and RBM are the slowest converging, particularly from higher energy. Modifying the RBM to be a FC-RBM  achieves similar energies at slightly lower epoch number. Both of these converge significantly slower than the FCNN and CNN architectures.

\subsection{Matching SWO} \label{subsec:match}
\begin{figure}
   \includegraphics[width=7cm]{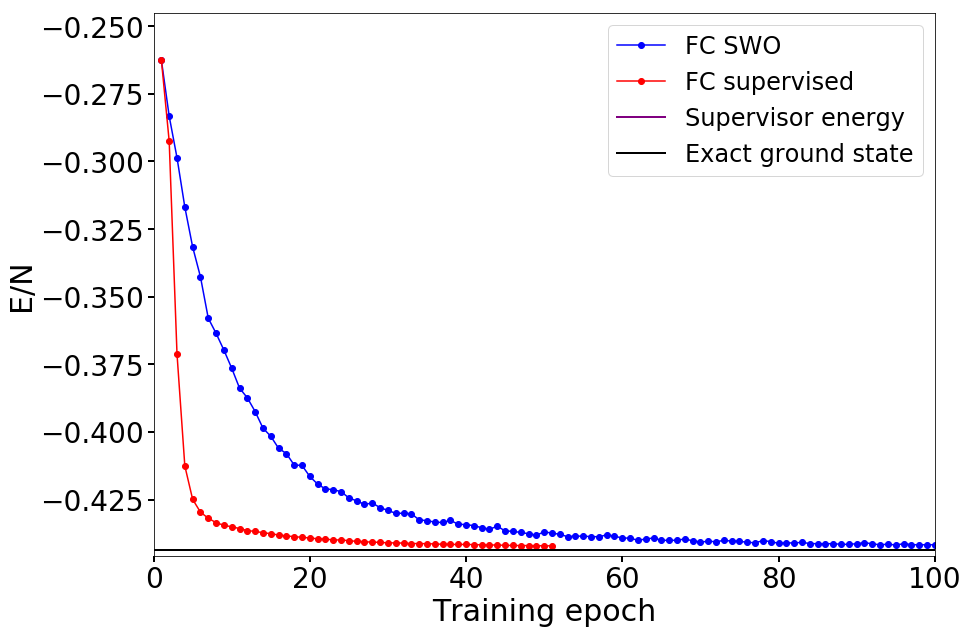}
   \caption{FCNN trained using IT-SWO (blue) and using a pre-trained CNN (red).  The pre-trained CNN has an energy shown by the purple line. Training the FCNN with the CNN supervisor converges significantly faster.}
   \label{fig:wavefunction_matching}
   \end{figure}

In addition to examples optimizing with IT-SWO, we also show how to use SWO to match a wave-function from an already optimized wave-function of a different architecture.  In particular, we first optimize a CNN architecture and then train a FCNN ansatz using the CNN as the superviser (see fig.~\ref{fig:wavefunction_matching}).   Notice training the FCNN using the converged CNN is much more efficient then training the FCNN directly via imaginary time evolution.  This approach then has at least two potential advantages.  First, as in this case, it might allow for more efficient convergence to the ground state by starting off with a quicker to optimize wave-function.  Secondly, one of the problems of optimization is local minima which can be induced by an inability to represent intermediate optimization states even if a more accurate final state is representable.  By matching with a different architecture it is potentially possible to overcome such problems. 

\subsection{System size agnostic architectures}

\begin{figure}
    \includegraphics[width=7cm]{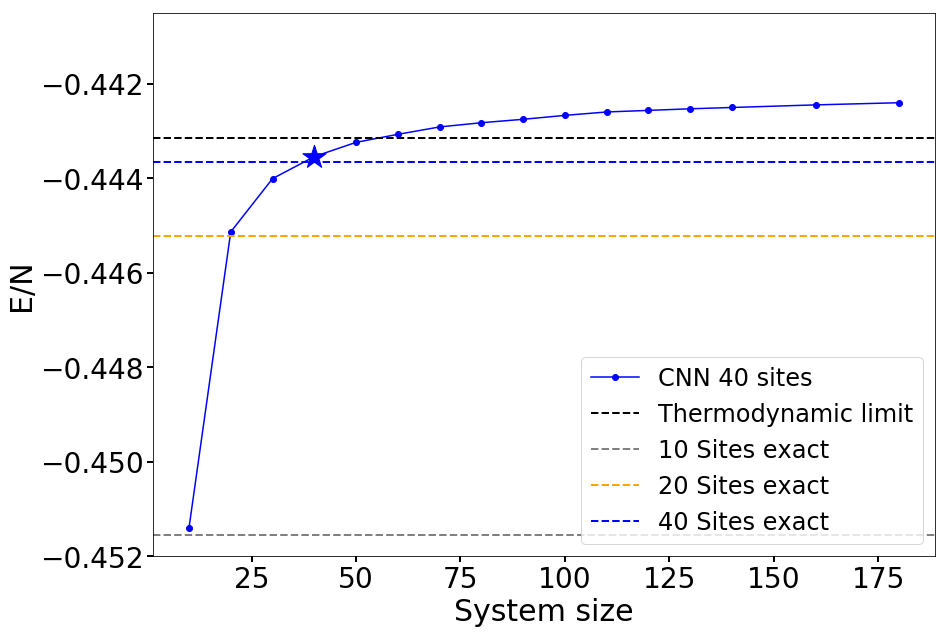}
  \caption{Evaluation of CNN ansatz on different system sizes. The weights on the wave-function are optimized by IT-SWO on a 40 sites Heisenberg chain and then evaluated on various system sizes using MCMC.}
  \label{fig:invariant_cnn}
\end{figure}

Among the architectures considered in the previous section, CNN and ResNet have the property that the parameters are optimized for a finite size kernel which is then applied over all the sites.  This means it is straightforward to take a kernel which is optimized on one system size and apply it to another system size using the same kernel size and parameters. This allows us to see how dependent the optimization is on system size. 

Specifically, we optimize the CNN on a 40-site 1D Heisenberg model and then consider the energy of this CNN with fixed weights on a range of system sizes (see fig~\ref{fig:invariant_cnn}).  Interestingly, not only does the energy per site remain consistently close to the true thermodynamic ground state answer for larger system sizes, but for smaller system sizes ($N\approx 20,30$) it also produces almost exact energies per site in spite of their relative far distance from the energy of the larger systems. Note that these energies are all variational upper bounds for the respective system sizes.

\section{Discussion}
\label{sec:Discussion}

The primary contributions of this work are (1) the introduction of a new class of variational wave-functions \textit{computational graph states}, that allow the flexible design of wave-functions based on their computational architecture and (2) the development of \textit{supervised wave-function optimization (SWO)}, a new optimization scheme to optimize wave-functions. While these approaches can be used separately there is non-trivial synergy in applying them together. 

In addition we have introduced a number of architectures for variational ansatz which haven't previously been considered, such as ResNet and  multi-RBM, as well as new variants of previously considered networks, that show promising performance on our benchmarks. This shows that, in spite of the current focus in the variational machine learning wave-functions community on restricted Boltzmann machines, there are a plethora of other architectures which achieve competitive results on similar problems.  

Of particular interest in this exploration is the discovery that the CNN architecture can be optimized on a single system size and give an efficient variational wave-function over a wide range of other system sizes.  

In the process of this study, we have developed an efficient code, CGS-VMC, (released at [https://github.com/ClarkResearchGroup/cgs-vmc]) which implements SWO and the computational graph framework. We anticipate it being useful to other researchers exploring alternative architectures.

The success of the variational approach to quantum mechanics depends on the representability of your variational ansatz and the optimization landscape under your optimization procedure.  We have now reached the point in variational wave-functions where the class of ansatz under consideration essentially encompasses all efficient computer programs.  It then behooves us to consider the best way to represent such computations.  Our answer to this problem is to represent these states as a CGS.   It is interesting to note that the variational representation of the full class of wave-functions which are efficient on a quantum computer already plays an important role in the quantum computing literature \cite{peruzzo2014variational}. 

The optimization landscape for CGS depends on both a discrete choice of architecture as well as a continuous optimization over the parameters within that architecture.  While we have not made progress on automating this discrete choice in this work, we have developed a formalism and approach which allows for  rapid prototyping and exploration over many different architectures.

In the long run, we will want automatically optimize over different CGS architectures and works in the machine learning community on automatic neural architecture search \cite{zoph2016neural,pham2018efficient} suggests a starting point toward accomplishing this. 

To make progress on the optimization of continuous parameters, we have developed SWO. The goal of any optimization technique is to find the state closest to the ground state and their primary obstacles are problems of local minima and slow convergence. IT-SWO is an efficient, simple first order method but follows the imaginary time propagator which is known to converge significantly faster than standard stochastic gradient descent. Additionally SWO naturally suggests a systematic way of avoiding poor local minimas. In the case of IT-SWO we can think of local minima as a phenomena arising from inability to represent some intermediate states along the imaginary time evolution trajectory. By training multiple architectures simultaneously, as long as one architecture can pass the barrier, it can serve as a supervisor to other architectures, potentially more suitable for approximating the ground state. Alternatively, there may be other ways of targeting better states which tunnel through barriers.  For example one might use a higher order Lanczos step \cite{sorella2001generalized}, a different propagator which has the ground state as a fixed point (such as  $\prod e^{J\hat{S}_{i} \hat{S}_{j}}$ for the Heisenberg model), or a stochastic sample of the ground state.  Such alternatives may affect convergence speeds as well.   
  
The traditional cycle of developing new variational ansatzs in the field of Variational Monte Carlo currently involves significant work in implementing both the variational ansatz as well as all the derivatives with respect to parameters greatly limiting the scope of models that are being considered and the rate at which various atomic and combined ansatz can be tested.  Working directly with CGS overcomes this;  architectures can be changed with minimal work and derivatives can be  automatically computed by automatic differentiation.  Combined with SWO, this approach allows for accelerated discovery of optimal wave-function ansatz to the quantum many-body problem.

\subsection*{Acknowledgements}
We thank Ryan Levy for the DMRG data in fig.~\ref{fig:1d_40_sums}.  We acknowledge useful conversations with Di Luo.

\bibliography{GraphSWO}

\begin{thebibliography}{64}%
\makeatletter
\providecommand \@ifxundefined [1]{%
 \@ifx{#1\undefined}
}%
\providecommand \@ifnum [1]{%
 \ifnum #1\expandafter \@firstoftwo
 \else \expandafter \@secondoftwo
 \fi
}%
\providecommand \@ifx [1]{%
 \ifx #1\expandafter \@firstoftwo
 \else \expandafter \@secondoftwo
 \fi
}%
\providecommand \natexlab [1]{#1}%
\providecommand \enquote  [1]{``#1''}%
\providecommand \bibnamefont  [1]{#1}%
\providecommand \bibfnamefont [1]{#1}%
\providecommand \citenamefont [1]{#1}%
\providecommand \href@noop [0]{\@secondoftwo}%
\providecommand \href [0]{\begingroup \@sanitize@url \@href}%
\providecommand \@href[1]{\@@startlink{#1}\@@href}%
\providecommand \@@href[1]{\endgroup#1\@@endlink}%
\providecommand \@sanitize@url [0]{\catcode `\\12\catcode `\$12\catcode
  `\&12\catcode `\#12\catcode `\^12\catcode `\_12\catcode `\%12\relax}%
\providecommand \@@startlink[1]{}%
\providecommand \@@endlink[0]{}%
\providecommand \url  [0]{\begingroup\@sanitize@url \@url }%
\providecommand \@url [1]{\endgroup\@href {#1}{\urlprefix }}%
\providecommand \urlprefix  [0]{URL }%
\providecommand \Eprint [0]{\href }%
\providecommand \doibase [0]{http://dx.doi.org/}%
\providecommand \selectlanguage [0]{\@gobble}%
\providecommand \bibinfo  [0]{\@secondoftwo}%
\providecommand \bibfield  [0]{\@secondoftwo}%
\providecommand \translation [1]{[#1]}%
\providecommand \BibitemOpen [0]{}%
\providecommand \bibitemStop [0]{}%
\providecommand \bibitemNoStop [0]{.\EOS\space}%
\providecommand \EOS [0]{\spacefactor3000\relax}%
\providecommand \BibitemShut  [1]{\csname bibitem#1\endcsname}%
\let\auto@bib@innerbib\@empty
\bibitem [{\citenamefont {McMillan}(1965)}]{mcmillan1965ground}%
  \BibitemOpen
  \bibfield  {author} {\bibinfo {author} {\bibfnamefont {W.~L.}\ \bibnamefont
  {McMillan}},\ }\href@noop {} {\bibfield  {journal} {\bibinfo  {journal}
  {Physical Review}\ }\textbf {\bibinfo {volume} {138}},\ \bibinfo {pages}
  {A442} (\bibinfo {year} {1965})}\BibitemShut {NoStop}%
\bibitem [{\citenamefont {Slater}(1951)}]{slater1951simplification}%
  \BibitemOpen
  \bibfield  {author} {\bibinfo {author} {\bibfnamefont {J.~C.}\ \bibnamefont
  {Slater}},\ }\href@noop {} {\bibfield  {journal} {\bibinfo  {journal}
  {Physical review}\ }\textbf {\bibinfo {volume} {81}},\ \bibinfo {pages} {385}
  (\bibinfo {year} {1951})}\BibitemShut {NoStop}%
\bibitem [{\citenamefont {Schollw{\"o}ck}(2011)}]{schollwock2011density}%
  \BibitemOpen
  \bibfield  {author} {\bibinfo {author} {\bibfnamefont {U.}~\bibnamefont
  {Schollw{\"o}ck}},\ }\href@noop {} {\bibfield  {journal} {\bibinfo  {journal}
  {Annals of Physics}\ }\textbf {\bibinfo {volume} {326}},\ \bibinfo {pages}
  {96} (\bibinfo {year} {2011})}\BibitemShut {NoStop}%
\bibitem [{\citenamefont {Vidal}(2003)}]{vidal2003efficient}%
  \BibitemOpen
  \bibfield  {author} {\bibinfo {author} {\bibfnamefont {G.}~\bibnamefont
  {Vidal}},\ }\href@noop {} {\bibfield  {journal} {\bibinfo  {journal}
  {Physical review letters}\ }\textbf {\bibinfo {volume} {91}},\ \bibinfo
  {pages} {147902} (\bibinfo {year} {2003})}\BibitemShut {NoStop}%
\bibitem [{\citenamefont {White}(1992)}]{white1992density}%
  \BibitemOpen
  \bibfield  {author} {\bibinfo {author} {\bibfnamefont {S.~R.}\ \bibnamefont
  {White}},\ }\href@noop {} {\bibfield  {journal} {\bibinfo  {journal}
  {Physical review letters}\ }\textbf {\bibinfo {volume} {69}},\ \bibinfo
  {pages} {2863} (\bibinfo {year} {1992})}\BibitemShut {NoStop}%
\bibitem [{\citenamefont {Capriotti}\ \emph {et~al.}(2002)\citenamefont
  {Capriotti}, \citenamefont {Becca}, \citenamefont {Parola},\ and\
  \citenamefont {Sorella}}]{capriotti2002projected}%
  \BibitemOpen
  \bibfield  {author} {\bibinfo {author} {\bibfnamefont {L.}~\bibnamefont
  {Capriotti}}, \bibinfo {author} {\bibfnamefont {F.}~\bibnamefont {Becca}},
  \bibinfo {author} {\bibfnamefont {A.}~\bibnamefont {Parola}}, \ and\ \bibinfo
  {author} {\bibfnamefont {S.}~\bibnamefont {Sorella}},\ }\href@noop {}
  {\bibfield  {journal} {\bibinfo  {journal} {arXiv preprint cond-mat/0208371}\
  } (\bibinfo {year} {2002})}\BibitemShut {NoStop}%
\bibitem [{\citenamefont {Huse}\ and\ \citenamefont
  {Elser}(1988)}]{huse1988simple}%
  \BibitemOpen
  \bibfield  {author} {\bibinfo {author} {\bibfnamefont {D.~A.}\ \bibnamefont
  {Huse}}\ and\ \bibinfo {author} {\bibfnamefont {V.}~\bibnamefont {Elser}},\
  }\href@noop {} {\bibfield  {journal} {\bibinfo  {journal} {Physical review
  letters}\ }\textbf {\bibinfo {volume} {60}},\ \bibinfo {pages} {2531}
  (\bibinfo {year} {1988})}\BibitemShut {NoStop}%
\bibitem [{\citenamefont {Changlani}\ \emph {et~al.}(2009)\citenamefont
  {Changlani}, \citenamefont {Kinder}, \citenamefont {Umrigar},\ and\
  \citenamefont {Chan}}]{changlani2009approximating}%
  \BibitemOpen
  \bibfield  {author} {\bibinfo {author} {\bibfnamefont {H.~J.}\ \bibnamefont
  {Changlani}}, \bibinfo {author} {\bibfnamefont {J.~M.}\ \bibnamefont
  {Kinder}}, \bibinfo {author} {\bibfnamefont {C.~J.}\ \bibnamefont {Umrigar}},
  \ and\ \bibinfo {author} {\bibfnamefont {G.~K.-L.}\ \bibnamefont {Chan}},\
  }\href@noop {} {\bibfield  {journal} {\bibinfo  {journal} {Physical Review
  B}\ }\textbf {\bibinfo {volume} {80}},\ \bibinfo {pages} {245116} (\bibinfo
  {year} {2009})}\BibitemShut {NoStop}%
\bibitem [{\citenamefont {Mezzacapo}\ \emph {et~al.}(2009)\citenamefont
  {Mezzacapo}, \citenamefont {Schuch}, \citenamefont {Boninsegni},\ and\
  \citenamefont {Cirac}}]{mezzacapo2009ground}%
  \BibitemOpen
  \bibfield  {author} {\bibinfo {author} {\bibfnamefont {F.}~\bibnamefont
  {Mezzacapo}}, \bibinfo {author} {\bibfnamefont {N.}~\bibnamefont {Schuch}},
  \bibinfo {author} {\bibfnamefont {M.}~\bibnamefont {Boninsegni}}, \ and\
  \bibinfo {author} {\bibfnamefont {J.~I.}\ \bibnamefont {Cirac}},\ }\href@noop
  {} {\bibfield  {journal} {\bibinfo  {journal} {New Journal of Physics}\
  }\textbf {\bibinfo {volume} {11}},\ \bibinfo {pages} {083026} (\bibinfo
  {year} {2009})}\BibitemShut {NoStop}%
\bibitem [{\citenamefont {Marti}\ \emph {et~al.}(2010)\citenamefont {Marti},
  \citenamefont {Bauer}, \citenamefont {Reiher}, \citenamefont {Troyer},\ and\
  \citenamefont {Verstraete}}]{marti2010complete}%
  \BibitemOpen
  \bibfield  {author} {\bibinfo {author} {\bibfnamefont {K.~H.}\ \bibnamefont
  {Marti}}, \bibinfo {author} {\bibfnamefont {B.}~\bibnamefont {Bauer}},
  \bibinfo {author} {\bibfnamefont {M.}~\bibnamefont {Reiher}}, \bibinfo
  {author} {\bibfnamefont {M.}~\bibnamefont {Troyer}}, \ and\ \bibinfo {author}
  {\bibfnamefont {F.}~\bibnamefont {Verstraete}},\ }\href@noop {} {\bibfield
  {journal} {\bibinfo  {journal} {New Journal of Physics}\ }\textbf {\bibinfo
  {volume} {12}},\ \bibinfo {pages} {103008} (\bibinfo {year}
  {2010})}\BibitemShut {NoStop}%
\bibitem [{\citenamefont {Carleo}\ and\ \citenamefont
  {Troyer}(2017)}]{carleo2017solving}%
  \BibitemOpen
  \bibfield  {author} {\bibinfo {author} {\bibfnamefont {G.}~\bibnamefont
  {Carleo}}\ and\ \bibinfo {author} {\bibfnamefont {M.}~\bibnamefont
  {Troyer}},\ }\href@noop {} {\bibfield  {journal} {\bibinfo  {journal}
  {Science}\ }\textbf {\bibinfo {volume} {355}},\ \bibinfo {pages} {602}
  (\bibinfo {year} {2017})}\BibitemShut {NoStop}%
\bibitem [{\citenamefont {Deng}\ \emph
  {et~al.}(2017{\natexlab{a}})\citenamefont {Deng}, \citenamefont {Li},\ and\
  \citenamefont {Sarma}}]{deng2017machine}%
  \BibitemOpen
  \bibfield  {author} {\bibinfo {author} {\bibfnamefont {D.-L.}\ \bibnamefont
  {Deng}}, \bibinfo {author} {\bibfnamefont {X.}~\bibnamefont {Li}}, \ and\
  \bibinfo {author} {\bibfnamefont {S.~D.}\ \bibnamefont {Sarma}},\ }\href@noop
  {} {\bibfield  {journal} {\bibinfo  {journal} {Physical Review B}\ }\textbf
  {\bibinfo {volume} {96}},\ \bibinfo {pages} {195145} (\bibinfo {year}
  {2017}{\natexlab{a}})}\BibitemShut {NoStop}%
\bibitem [{\citenamefont {Kaubruegger}\ \emph {et~al.}(2018)\citenamefont
  {Kaubruegger}, \citenamefont {Pastori},\ and\ \citenamefont
  {Budich}}]{kaubruegger2018chiral}%
  \BibitemOpen
  \bibfield  {author} {\bibinfo {author} {\bibfnamefont {R.}~\bibnamefont
  {Kaubruegger}}, \bibinfo {author} {\bibfnamefont {L.}~\bibnamefont
  {Pastori}}, \ and\ \bibinfo {author} {\bibfnamefont {J.~C.}\ \bibnamefont
  {Budich}},\ }\href@noop {} {\bibfield  {journal} {\bibinfo  {journal}
  {Physical Review B}\ }\textbf {\bibinfo {volume} {97}},\ \bibinfo {pages}
  {195136} (\bibinfo {year} {2018})}\BibitemShut {NoStop}%
\bibitem [{\citenamefont {Torlai}\ \emph {et~al.}(2018)\citenamefont {Torlai},
  \citenamefont {Mazzola}, \citenamefont {Carrasquilla}, \citenamefont
  {Troyer}, \citenamefont {Melko},\ and\ \citenamefont
  {Carleo}}]{torlai2018neural}%
  \BibitemOpen
  \bibfield  {author} {\bibinfo {author} {\bibfnamefont {G.}~\bibnamefont
  {Torlai}}, \bibinfo {author} {\bibfnamefont {G.}~\bibnamefont {Mazzola}},
  \bibinfo {author} {\bibfnamefont {J.}~\bibnamefont {Carrasquilla}}, \bibinfo
  {author} {\bibfnamefont {M.}~\bibnamefont {Troyer}}, \bibinfo {author}
  {\bibfnamefont {R.}~\bibnamefont {Melko}}, \ and\ \bibinfo {author}
  {\bibfnamefont {G.}~\bibnamefont {Carleo}},\ }\href@noop {} {\bibfield
  {journal} {\bibinfo  {journal} {Nature Physics}\ }\textbf {\bibinfo {volume}
  {14}},\ \bibinfo {pages} {447} (\bibinfo {year} {2018})}\BibitemShut
  {NoStop}%
\bibitem [{\citenamefont {Deng}\ \emph
  {et~al.}(2017{\natexlab{b}})\citenamefont {Deng}, \citenamefont {Li},\ and\
  \citenamefont {Sarma}}]{deng2017quantum}%
  \BibitemOpen
  \bibfield  {author} {\bibinfo {author} {\bibfnamefont {D.-L.}\ \bibnamefont
  {Deng}}, \bibinfo {author} {\bibfnamefont {X.}~\bibnamefont {Li}}, \ and\
  \bibinfo {author} {\bibfnamefont {S.~D.}\ \bibnamefont {Sarma}},\ }\href@noop
  {} {\bibfield  {journal} {\bibinfo  {journal} {Physical Review X}\ }\textbf
  {\bibinfo {volume} {7}},\ \bibinfo {pages} {021021} (\bibinfo {year}
  {2017}{\natexlab{b}})}\BibitemShut {NoStop}%
\bibitem [{\citenamefont {Glasser}\ \emph {et~al.}(2018)\citenamefont
  {Glasser}, \citenamefont {Pancotti}, \citenamefont {August}, \citenamefont
  {Rodriguez},\ and\ \citenamefont {Cirac}}]{glasser2018neural}%
  \BibitemOpen
  \bibfield  {author} {\bibinfo {author} {\bibfnamefont {I.}~\bibnamefont
  {Glasser}}, \bibinfo {author} {\bibfnamefont {N.}~\bibnamefont {Pancotti}},
  \bibinfo {author} {\bibfnamefont {M.}~\bibnamefont {August}}, \bibinfo
  {author} {\bibfnamefont {I.~D.}\ \bibnamefont {Rodriguez}}, \ and\ \bibinfo
  {author} {\bibfnamefont {J.~I.}\ \bibnamefont {Cirac}},\ }\href@noop {}
  {\bibfield  {journal} {\bibinfo  {journal} {Physical Review X}\ }\textbf
  {\bibinfo {volume} {8}},\ \bibinfo {pages} {011006} (\bibinfo {year}
  {2018})}\BibitemShut {NoStop}%
\bibitem [{\citenamefont {Choo}\ \emph {et~al.}(2018)\citenamefont {Choo},
  \citenamefont {Carleo}, \citenamefont {Regnault},\ and\ \citenamefont
  {Neupert}}]{choo2018symmetries}%
  \BibitemOpen
  \bibfield  {author} {\bibinfo {author} {\bibfnamefont {K.}~\bibnamefont
  {Choo}}, \bibinfo {author} {\bibfnamefont {G.}~\bibnamefont {Carleo}},
  \bibinfo {author} {\bibfnamefont {N.}~\bibnamefont {Regnault}}, \ and\
  \bibinfo {author} {\bibfnamefont {T.}~\bibnamefont {Neupert}},\ }\href@noop
  {} {\bibfield  {journal} {\bibinfo  {journal} {Physical review letters}\
  }\textbf {\bibinfo {volume} {121}},\ \bibinfo {pages} {167204} (\bibinfo
  {year} {2018})}\BibitemShut {NoStop}%
\bibitem [{\citenamefont {Martin}\ \emph {et~al.}(2016)\citenamefont {Martin},
  \citenamefont {Reining},\ and\ \citenamefont
  {Ceperley}}]{martin2016interacting}%
  \BibitemOpen
  \bibfield  {author} {\bibinfo {author} {\bibfnamefont {R.~M.}\ \bibnamefont
  {Martin}}, \bibinfo {author} {\bibfnamefont {L.}~\bibnamefont {Reining}}, \
  and\ \bibinfo {author} {\bibfnamefont {D.~M.}\ \bibnamefont {Ceperley}},\
  }\href@noop {} {\emph {\bibinfo {title} {Interacting electrons}}}\ (\bibinfo
  {publisher} {Cambridge University Press},\ \bibinfo {year}
  {2016})\BibitemShut {NoStop}%
\bibitem [{\citenamefont {Chou}\ \emph {et~al.}(2012)\citenamefont {Chou},
  \citenamefont {Pollmann},\ and\ \citenamefont {Lee}}]{chou2012matrix}%
  \BibitemOpen
  \bibfield  {author} {\bibinfo {author} {\bibfnamefont {C.-P.}\ \bibnamefont
  {Chou}}, \bibinfo {author} {\bibfnamefont {F.}~\bibnamefont {Pollmann}}, \
  and\ \bibinfo {author} {\bibfnamefont {T.-K.}\ \bibnamefont {Lee}},\
  }\href@noop {} {\bibfield  {journal} {\bibinfo  {journal} {Physical Review
  B}\ }\textbf {\bibinfo {volume} {86}},\ \bibinfo {pages} {041105} (\bibinfo
  {year} {2012})}\BibitemShut {NoStop}%
\bibitem [{\citenamefont {Nomura}\ \emph {et~al.}(2017)\citenamefont {Nomura},
  \citenamefont {Darmawan}, \citenamefont {Yamaji},\ and\ \citenamefont
  {Imada}}]{nomura2017restricted}%
  \BibitemOpen
  \bibfield  {author} {\bibinfo {author} {\bibfnamefont {Y.}~\bibnamefont
  {Nomura}}, \bibinfo {author} {\bibfnamefont {A.~S.}\ \bibnamefont
  {Darmawan}}, \bibinfo {author} {\bibfnamefont {Y.}~\bibnamefont {Yamaji}}, \
  and\ \bibinfo {author} {\bibfnamefont {M.}~\bibnamefont {Imada}},\
  }\href@noop {} {\bibfield  {journal} {\bibinfo  {journal} {Physical Review
  B}\ }\textbf {\bibinfo {volume} {96}},\ \bibinfo {pages} {205152} (\bibinfo
  {year} {2017})}\BibitemShut {NoStop}%
\bibitem [{\citenamefont {Olsen}\ \emph {et~al.}(1988)\citenamefont {Olsen},
  \citenamefont {Roos}, \citenamefont {Jo/rgensen},\ and\ \citenamefont
  {Jensen}}]{olsen1988determinant}%
  \BibitemOpen
  \bibfield  {author} {\bibinfo {author} {\bibfnamefont {J.}~\bibnamefont
  {Olsen}}, \bibinfo {author} {\bibfnamefont {B.~O.}\ \bibnamefont {Roos}},
  \bibinfo {author} {\bibfnamefont {P.}~\bibnamefont {Jo/rgensen}}, \ and\
  \bibinfo {author} {\bibfnamefont {H.~J.~A.}\ \bibnamefont {Jensen}},\
  }\href@noop {} {\bibfield  {journal} {\bibinfo  {journal} {The Journal of
  chemical physics}\ }\textbf {\bibinfo {volume} {89}},\ \bibinfo {pages}
  {2185} (\bibinfo {year} {1988})}\BibitemShut {NoStop}%
\bibitem [{\citenamefont {Holmes}\ \emph {et~al.}(2016)\citenamefont {Holmes},
  \citenamefont {Tubman},\ and\ \citenamefont {Umrigar}}]{holmes2016heat}%
  \BibitemOpen
  \bibfield  {author} {\bibinfo {author} {\bibfnamefont {A.~A.}\ \bibnamefont
  {Holmes}}, \bibinfo {author} {\bibfnamefont {N.~M.}\ \bibnamefont {Tubman}},
  \ and\ \bibinfo {author} {\bibfnamefont {C.}~\bibnamefont {Umrigar}},\
  }\href@noop {} {\bibfield  {journal} {\bibinfo  {journal} {Journal of
  chemical theory and computation}\ }\textbf {\bibinfo {volume} {12}},\
  \bibinfo {pages} {3674} (\bibinfo {year} {2016})}\BibitemShut {NoStop}%
\bibitem [{\citenamefont {Chang}\ and\ \citenamefont
  {Morales}(2017)}]{chang2017multi}%
  \BibitemOpen
  \bibfield  {author} {\bibinfo {author} {\bibfnamefont {C.-C.}\ \bibnamefont
  {Chang}}\ and\ \bibinfo {author} {\bibfnamefont {M.~A.}\ \bibnamefont
  {Morales}},\ }\href@noop {} {\bibfield  {journal} {\bibinfo  {journal} {arXiv
  preprint arXiv:1711.02154}\ } (\bibinfo {year} {2017})}\BibitemShut {NoStop}%
\bibitem [{\citenamefont {Bouab{\c{c}}a}\ \emph {et~al.}(2010)\citenamefont
  {Bouab{\c{c}}a}, \citenamefont {Bra{\"\i}da},\ and\ \citenamefont
  {Caffarel}}]{bouabcca2010multi}%
  \BibitemOpen
  \bibfield  {author} {\bibinfo {author} {\bibfnamefont {T.}~\bibnamefont
  {Bouab{\c{c}}a}}, \bibinfo {author} {\bibfnamefont {B.}~\bibnamefont
  {Bra{\"\i}da}}, \ and\ \bibinfo {author} {\bibfnamefont {M.}~\bibnamefont
  {Caffarel}},\ }\href@noop {} {\bibfield  {journal} {\bibinfo  {journal} {The
  Journal of chemical physics}\ }\textbf {\bibinfo {volume} {133}},\ \bibinfo
  {pages} {044111} (\bibinfo {year} {2010})}\BibitemShut {NoStop}%
\bibitem [{\citenamefont {Morales}\ \emph {et~al.}(2012)\citenamefont
  {Morales}, \citenamefont {McMinis}, \citenamefont {Clark}, \citenamefont
  {Kim},\ and\ \citenamefont {Scuseria}}]{morales2012multideterminant}%
  \BibitemOpen
  \bibfield  {author} {\bibinfo {author} {\bibfnamefont {M.~A.}\ \bibnamefont
  {Morales}}, \bibinfo {author} {\bibfnamefont {J.}~\bibnamefont {McMinis}},
  \bibinfo {author} {\bibfnamefont {B.~K.}\ \bibnamefont {Clark}}, \bibinfo
  {author} {\bibfnamefont {J.}~\bibnamefont {Kim}}, \ and\ \bibinfo {author}
  {\bibfnamefont {G.~E.}\ \bibnamefont {Scuseria}},\ }\href@noop {} {\bibfield
  {journal} {\bibinfo  {journal} {Journal of chemical theory and computation}\
  }\textbf {\bibinfo {volume} {8}},\ \bibinfo {pages} {2181} (\bibinfo {year}
  {2012})}\BibitemShut {NoStop}%
\bibitem [{\citenamefont {Luo}\ and\ \citenamefont
  {Clark}(2018)}]{luo2018backflow}%
  \BibitemOpen
  \bibfield  {author} {\bibinfo {author} {\bibfnamefont {D.}~\bibnamefont
  {Luo}}\ and\ \bibinfo {author} {\bibfnamefont {B.~K.}\ \bibnamefont
  {Clark}},\ }\href@noop {} {\bibfield  {journal} {\bibinfo  {journal} {arXiv
  preprint arXiv:1807.10770}\ } (\bibinfo {year} {2018})}\BibitemShut {NoStop}%
\bibitem [{\citenamefont {Taddei}\ \emph {et~al.}(2015)\citenamefont {Taddei},
  \citenamefont {Ruggeri}, \citenamefont {Moroni},\ and\ \citenamefont
  {Holzmann}}]{taddei2015iterative}%
  \BibitemOpen
  \bibfield  {author} {\bibinfo {author} {\bibfnamefont {M.}~\bibnamefont
  {Taddei}}, \bibinfo {author} {\bibfnamefont {M.}~\bibnamefont {Ruggeri}},
  \bibinfo {author} {\bibfnamefont {S.}~\bibnamefont {Moroni}}, \ and\ \bibinfo
  {author} {\bibfnamefont {M.}~\bibnamefont {Holzmann}},\ }\href@noop {}
  {\bibfield  {journal} {\bibinfo  {journal} {Physical Review B}\ }\textbf
  {\bibinfo {volume} {91}},\ \bibinfo {pages} {115106} (\bibinfo {year}
  {2015})}\BibitemShut {NoStop}%
\bibitem [{\citenamefont {Feynman}\ and\ \citenamefont
  {Cohen}(1956)}]{feynman1956energy}%
  \BibitemOpen
  \bibfield  {author} {\bibinfo {author} {\bibfnamefont {R.}~\bibnamefont
  {Feynman}}\ and\ \bibinfo {author} {\bibfnamefont {M.}~\bibnamefont
  {Cohen}},\ }\href@noop {} {\bibfield  {journal} {\bibinfo  {journal}
  {Physical Review}\ }\textbf {\bibinfo {volume} {102}},\ \bibinfo {pages}
  {1189} (\bibinfo {year} {1956})}\BibitemShut {NoStop}%
\bibitem [{\citenamefont {Abadi}\ \emph {et~al.}(2016)\citenamefont {Abadi},
  \citenamefont {Barham}, \citenamefont {Chen}, \citenamefont {Chen},
  \citenamefont {Davis}, \citenamefont {Dean}, \citenamefont {Devin},
  \citenamefont {Ghemawat}, \citenamefont {Irving}, \citenamefont {Isard} \emph
  {et~al.}}]{abadi2016tensorflow}%
  \BibitemOpen
  \bibfield  {author} {\bibinfo {author} {\bibfnamefont {M.}~\bibnamefont
  {Abadi}}, \bibinfo {author} {\bibfnamefont {P.}~\bibnamefont {Barham}},
  \bibinfo {author} {\bibfnamefont {J.}~\bibnamefont {Chen}}, \bibinfo {author}
  {\bibfnamefont {Z.}~\bibnamefont {Chen}}, \bibinfo {author} {\bibfnamefont
  {A.}~\bibnamefont {Davis}}, \bibinfo {author} {\bibfnamefont
  {J.}~\bibnamefont {Dean}}, \bibinfo {author} {\bibfnamefont {M.}~\bibnamefont
  {Devin}}, \bibinfo {author} {\bibfnamefont {S.}~\bibnamefont {Ghemawat}},
  \bibinfo {author} {\bibfnamefont {G.}~\bibnamefont {Irving}}, \bibinfo
  {author} {\bibfnamefont {M.}~\bibnamefont {Isard}},  \emph {et~al.},\ }in\
  \href@noop {} {\emph {\bibinfo {booktitle} {OSDI}}},\ Vol.~\bibinfo {volume}
  {16}\ (\bibinfo {year} {2016})\ pp.\ \bibinfo {pages} {265--283}\BibitemShut
  {NoStop}%
\bibitem [{\citenamefont {Al-Rfou}\ \emph {et~al.}(2016)\citenamefont
  {Al-Rfou}, \citenamefont {Alain}, \citenamefont {Almahairi}, \citenamefont
  {Angermueller}, \citenamefont {Bahdanau}, \citenamefont {Ballas},
  \citenamefont {Bastien}, \citenamefont {Bayer}, \citenamefont {Belikov},
  \citenamefont {Belopolsky} \emph {et~al.}}]{al2016theano}%
  \BibitemOpen
  \bibfield  {author} {\bibinfo {author} {\bibfnamefont {R.}~\bibnamefont
  {Al-Rfou}}, \bibinfo {author} {\bibfnamefont {G.}~\bibnamefont {Alain}},
  \bibinfo {author} {\bibfnamefont {A.}~\bibnamefont {Almahairi}}, \bibinfo
  {author} {\bibfnamefont {C.}~\bibnamefont {Angermueller}}, \bibinfo {author}
  {\bibfnamefont {D.}~\bibnamefont {Bahdanau}}, \bibinfo {author}
  {\bibfnamefont {N.}~\bibnamefont {Ballas}}, \bibinfo {author} {\bibfnamefont
  {F.}~\bibnamefont {Bastien}}, \bibinfo {author} {\bibfnamefont
  {J.}~\bibnamefont {Bayer}}, \bibinfo {author} {\bibfnamefont
  {A.}~\bibnamefont {Belikov}}, \bibinfo {author} {\bibfnamefont
  {A.}~\bibnamefont {Belopolsky}},  \emph {et~al.},\ }\href@noop {} {\bibfield
  {journal} {\bibinfo  {journal} {arXiv preprint}\ } (\bibinfo {year}
  {2016})}\BibitemShut {NoStop}%
\bibitem [{\citenamefont {Akidau}\ \emph {et~al.}(2015)\citenamefont {Akidau},
  \citenamefont {Bradshaw}, \citenamefont {Chambers}, \citenamefont {Chernyak},
  \citenamefont {Fern{\'a}ndez-Moctezuma}, \citenamefont {Lax}, \citenamefont
  {McVeety}, \citenamefont {Mills}, \citenamefont {Perry}, \citenamefont
  {Schmidt} \emph {et~al.}}]{akidau2015dataflow}%
  \BibitemOpen
  \bibfield  {author} {\bibinfo {author} {\bibfnamefont {T.}~\bibnamefont
  {Akidau}}, \bibinfo {author} {\bibfnamefont {R.}~\bibnamefont {Bradshaw}},
  \bibinfo {author} {\bibfnamefont {C.}~\bibnamefont {Chambers}}, \bibinfo
  {author} {\bibfnamefont {S.}~\bibnamefont {Chernyak}}, \bibinfo {author}
  {\bibfnamefont {R.~J.}\ \bibnamefont {Fern{\'a}ndez-Moctezuma}}, \bibinfo
  {author} {\bibfnamefont {R.}~\bibnamefont {Lax}}, \bibinfo {author}
  {\bibfnamefont {S.}~\bibnamefont {McVeety}}, \bibinfo {author} {\bibfnamefont
  {D.}~\bibnamefont {Mills}}, \bibinfo {author} {\bibfnamefont
  {F.}~\bibnamefont {Perry}}, \bibinfo {author} {\bibfnamefont
  {E.}~\bibnamefont {Schmidt}},  \emph {et~al.},\ }\href@noop {} {\bibfield
  {journal} {\bibinfo  {journal} {Proceedings of the VLDB Endowment}\ }\textbf
  {\bibinfo {volume} {8}},\ \bibinfo {pages} {1792} (\bibinfo {year}
  {2015})}\BibitemShut {NoStop}%
\bibitem [{\citenamefont {Jia}\ \emph {et~al.}(2014)\citenamefont {Jia},
  \citenamefont {Shelhamer}, \citenamefont {Donahue}, \citenamefont {Karayev},
  \citenamefont {Long}, \citenamefont {Girshick}, \citenamefont {Guadarrama},\
  and\ \citenamefont {Darrell}}]{jia2014caffe}%
  \BibitemOpen
  \bibfield  {author} {\bibinfo {author} {\bibfnamefont {Y.}~\bibnamefont
  {Jia}}, \bibinfo {author} {\bibfnamefont {E.}~\bibnamefont {Shelhamer}},
  \bibinfo {author} {\bibfnamefont {J.}~\bibnamefont {Donahue}}, \bibinfo
  {author} {\bibfnamefont {S.}~\bibnamefont {Karayev}}, \bibinfo {author}
  {\bibfnamefont {J.}~\bibnamefont {Long}}, \bibinfo {author} {\bibfnamefont
  {R.}~\bibnamefont {Girshick}}, \bibinfo {author} {\bibfnamefont
  {S.}~\bibnamefont {Guadarrama}}, \ and\ \bibinfo {author} {\bibfnamefont
  {T.}~\bibnamefont {Darrell}},\ }\href@noop {} {\bibfield  {journal} {\bibinfo
   {journal} {arXiv preprint arXiv:1408.5093}\ } (\bibinfo {year}
  {2014})}\BibitemShut {NoStop}%
\bibitem [{\citenamefont {Seide}\ and\ \citenamefont
  {Agarwal}(2016)}]{seide2016cntk}%
  \BibitemOpen
  \bibfield  {author} {\bibinfo {author} {\bibfnamefont {F.}~\bibnamefont
  {Seide}}\ and\ \bibinfo {author} {\bibfnamefont {A.}~\bibnamefont
  {Agarwal}},\ }in\ \href@noop {} {\emph {\bibinfo {booktitle} {Proceedings of
  the 22nd ACM SIGKDD International Conference on Knowledge Discovery and Data
  Mining}}}\ (\bibinfo {organization} {ACM},\ \bibinfo {year} {2016})\ pp.\
  \bibinfo {pages} {2135--2135}\BibitemShut {NoStop}%
\bibitem [{\citenamefont {Sorella}\ and\ \citenamefont
  {Capriotti}(2010)}]{sorella2010algorithmic}%
  \BibitemOpen
  \bibfield  {author} {\bibinfo {author} {\bibfnamefont {S.}~\bibnamefont
  {Sorella}}\ and\ \bibinfo {author} {\bibfnamefont {L.}~\bibnamefont
  {Capriotti}},\ }\href@noop {} {\bibfield  {journal} {\bibinfo  {journal} {The
  Journal of chemical physics}\ }\textbf {\bibinfo {volume} {133}},\ \bibinfo
  {pages} {234111} (\bibinfo {year} {2010})}\BibitemShut {NoStop}%
\bibitem [{\citenamefont {LeCun}\ \emph {et~al.}(1990)\citenamefont {LeCun},
  \citenamefont {Boser}, \citenamefont {Denker}, \citenamefont {Henderson},
  \citenamefont {Howard}, \citenamefont {Hubbard},\ and\ \citenamefont
  {Jackel}}]{lecun1990handwritten}%
  \BibitemOpen
  \bibfield  {author} {\bibinfo {author} {\bibfnamefont {Y.}~\bibnamefont
  {LeCun}}, \bibinfo {author} {\bibfnamefont {B.~E.}\ \bibnamefont {Boser}},
  \bibinfo {author} {\bibfnamefont {J.~S.}\ \bibnamefont {Denker}}, \bibinfo
  {author} {\bibfnamefont {D.}~\bibnamefont {Henderson}}, \bibinfo {author}
  {\bibfnamefont {R.~E.}\ \bibnamefont {Howard}}, \bibinfo {author}
  {\bibfnamefont {W.~E.}\ \bibnamefont {Hubbard}}, \ and\ \bibinfo {author}
  {\bibfnamefont {L.~D.}\ \bibnamefont {Jackel}},\ }in\ \href@noop {} {\emph
  {\bibinfo {booktitle} {Advances in neural information processing systems}}}\
  (\bibinfo {year} {1990})\ pp.\ \bibinfo {pages} {396--404}\BibitemShut
  {NoStop}%
\bibitem [{\citenamefont {Hochreiter}\ and\ \citenamefont
  {Schmidhuber}(1997)}]{hochreiter1997long}%
  \BibitemOpen
  \bibfield  {author} {\bibinfo {author} {\bibfnamefont {S.}~\bibnamefont
  {Hochreiter}}\ and\ \bibinfo {author} {\bibfnamefont {J.}~\bibnamefont
  {Schmidhuber}},\ }\href@noop {} {\bibfield  {journal} {\bibinfo  {journal}
  {Neural computation}\ }\textbf {\bibinfo {volume} {9}},\ \bibinfo {pages}
  {1735} (\bibinfo {year} {1997})}\BibitemShut {NoStop}%
\bibitem [{\citenamefont {He}\ \emph {et~al.}(2016)\citenamefont {He},
  \citenamefont {Zhang}, \citenamefont {Ren},\ and\ \citenamefont
  {Sun}}]{he2016deep}%
  \BibitemOpen
  \bibfield  {author} {\bibinfo {author} {\bibfnamefont {K.}~\bibnamefont
  {He}}, \bibinfo {author} {\bibfnamefont {X.}~\bibnamefont {Zhang}}, \bibinfo
  {author} {\bibfnamefont {S.}~\bibnamefont {Ren}}, \ and\ \bibinfo {author}
  {\bibfnamefont {J.}~\bibnamefont {Sun}},\ }in\ \href@noop {} {\emph {\bibinfo
  {booktitle} {Proceedings of the IEEE conference on computer vision and
  pattern recognition}}}\ (\bibinfo {year} {2016})\ pp.\ \bibinfo {pages}
  {770--778}\BibitemShut {NoStop}%
\bibitem [{\citenamefont {Szegedy}\ \emph {et~al.}(2015)\citenamefont
  {Szegedy}, \citenamefont {Liu}, \citenamefont {Jia}, \citenamefont
  {Sermanet}, \citenamefont {Reed}, \citenamefont {Anguelov}, \citenamefont
  {Erhan}, \citenamefont {Vanhoucke},\ and\ \citenamefont
  {Rabinovich}}]{szegedy2015going}%
  \BibitemOpen
  \bibfield  {author} {\bibinfo {author} {\bibfnamefont {C.}~\bibnamefont
  {Szegedy}}, \bibinfo {author} {\bibfnamefont {W.}~\bibnamefont {Liu}},
  \bibinfo {author} {\bibfnamefont {Y.}~\bibnamefont {Jia}}, \bibinfo {author}
  {\bibfnamefont {P.}~\bibnamefont {Sermanet}}, \bibinfo {author}
  {\bibfnamefont {S.}~\bibnamefont {Reed}}, \bibinfo {author} {\bibfnamefont
  {D.}~\bibnamefont {Anguelov}}, \bibinfo {author} {\bibfnamefont
  {D.}~\bibnamefont {Erhan}}, \bibinfo {author} {\bibfnamefont
  {V.}~\bibnamefont {Vanhoucke}}, \ and\ \bibinfo {author} {\bibfnamefont
  {A.}~\bibnamefont {Rabinovich}},\ }in\ \href@noop {} {\emph {\bibinfo
  {booktitle} {Proceedings of the IEEE conference on computer vision and
  pattern recognition}}}\ (\bibinfo {year} {2015})\ pp.\ \bibinfo {pages}
  {1--9}\BibitemShut {NoStop}%
\bibitem [{\citenamefont {Defferrard}\ \emph {et~al.}(2016)\citenamefont
  {Defferrard}, \citenamefont {Bresson},\ and\ \citenamefont
  {Vandergheynst}}]{defferrard2016convolutional}%
  \BibitemOpen
  \bibfield  {author} {\bibinfo {author} {\bibfnamefont {M.}~\bibnamefont
  {Defferrard}}, \bibinfo {author} {\bibfnamefont {X.}~\bibnamefont {Bresson}},
  \ and\ \bibinfo {author} {\bibfnamefont {P.}~\bibnamefont {Vandergheynst}},\
  }in\ \href@noop {} {\emph {\bibinfo {booktitle} {Advances in Neural
  Information Processing Systems}}}\ (\bibinfo {year} {2016})\ pp.\ \bibinfo
  {pages} {3844--3852}\BibitemShut {NoStop}%
\bibitem [{\citenamefont {Krizhevsky}\ \emph {et~al.}(2012)\citenamefont
  {Krizhevsky}, \citenamefont {Sutskever},\ and\ \citenamefont
  {Hinton}}]{krizhevsky2012imagenet}%
  \BibitemOpen
  \bibfield  {author} {\bibinfo {author} {\bibfnamefont {A.}~\bibnamefont
  {Krizhevsky}}, \bibinfo {author} {\bibfnamefont {I.}~\bibnamefont
  {Sutskever}}, \ and\ \bibinfo {author} {\bibfnamefont {G.~E.}\ \bibnamefont
  {Hinton}},\ }in\ \href@noop {} {\emph {\bibinfo {booktitle} {Advances in
  neural information processing systems}}}\ (\bibinfo {year} {2012})\ pp.\
  \bibinfo {pages} {1097--1105}\BibitemShut {NoStop}%
\bibitem [{\citenamefont {Graves}\ \emph {et~al.}(2016)\citenamefont {Graves},
  \citenamefont {Wayne}, \citenamefont {Reynolds}, \citenamefont {Harley},
  \citenamefont {Danihelka}, \citenamefont {Grabska-Barwi{\'n}ska},
  \citenamefont {Colmenarejo}, \citenamefont {Grefenstette}, \citenamefont
  {Ramalho}, \citenamefont {Agapiou} \emph {et~al.}}]{graves2016hybrid}%
  \BibitemOpen
  \bibfield  {author} {\bibinfo {author} {\bibfnamefont {A.}~\bibnamefont
  {Graves}}, \bibinfo {author} {\bibfnamefont {G.}~\bibnamefont {Wayne}},
  \bibinfo {author} {\bibfnamefont {M.}~\bibnamefont {Reynolds}}, \bibinfo
  {author} {\bibfnamefont {T.}~\bibnamefont {Harley}}, \bibinfo {author}
  {\bibfnamefont {I.}~\bibnamefont {Danihelka}}, \bibinfo {author}
  {\bibfnamefont {A.}~\bibnamefont {Grabska-Barwi{\'n}ska}}, \bibinfo {author}
  {\bibfnamefont {S.~G.}\ \bibnamefont {Colmenarejo}}, \bibinfo {author}
  {\bibfnamefont {E.}~\bibnamefont {Grefenstette}}, \bibinfo {author}
  {\bibfnamefont {T.}~\bibnamefont {Ramalho}}, \bibinfo {author} {\bibfnamefont
  {J.}~\bibnamefont {Agapiou}},  \emph {et~al.},\ }\href@noop {} {\bibfield
  {journal} {\bibinfo  {journal} {Nature}\ }\textbf {\bibinfo {volume} {538}},\
  \bibinfo {pages} {471} (\bibinfo {year} {2016})}\BibitemShut {NoStop}%
\bibitem [{\citenamefont {Van Den~Oord}\ \emph {et~al.}(2016)\citenamefont {Van
  Den~Oord}, \citenamefont {Dieleman}, \citenamefont {Zen}, \citenamefont
  {Simonyan}, \citenamefont {Vinyals}, \citenamefont {Graves}, \citenamefont
  {Kalchbrenner}, \citenamefont {Senior},\ and\ \citenamefont
  {Kavukcuoglu}}]{van2016wavenet}%
  \BibitemOpen
  \bibfield  {author} {\bibinfo {author} {\bibfnamefont {A.}~\bibnamefont {Van
  Den~Oord}}, \bibinfo {author} {\bibfnamefont {S.}~\bibnamefont {Dieleman}},
  \bibinfo {author} {\bibfnamefont {H.}~\bibnamefont {Zen}}, \bibinfo {author}
  {\bibfnamefont {K.}~\bibnamefont {Simonyan}}, \bibinfo {author}
  {\bibfnamefont {O.}~\bibnamefont {Vinyals}}, \bibinfo {author} {\bibfnamefont
  {A.}~\bibnamefont {Graves}}, \bibinfo {author} {\bibfnamefont
  {N.}~\bibnamefont {Kalchbrenner}}, \bibinfo {author} {\bibfnamefont {A.~W.}\
  \bibnamefont {Senior}}, \ and\ \bibinfo {author} {\bibfnamefont
  {K.}~\bibnamefont {Kavukcuoglu}},\ }in\ \href@noop {} {\emph {\bibinfo
  {booktitle} {SSW}}}\ (\bibinfo {year} {2016})\ p.\ \bibinfo {pages}
  {125}\BibitemShut {NoStop}%
\bibitem [{\citenamefont {Bar-Sinai}\ \emph {et~al.}(2018)\citenamefont
  {Bar-Sinai}, \citenamefont {Hoyer}, \citenamefont {Hickey},\ and\
  \citenamefont {Brenner}}]{bar2018data}%
  \BibitemOpen
  \bibfield  {author} {\bibinfo {author} {\bibfnamefont {Y.}~\bibnamefont
  {Bar-Sinai}}, \bibinfo {author} {\bibfnamefont {S.}~\bibnamefont {Hoyer}},
  \bibinfo {author} {\bibfnamefont {J.}~\bibnamefont {Hickey}}, \ and\ \bibinfo
  {author} {\bibfnamefont {M.~P.}\ \bibnamefont {Brenner}},\ }\href@noop {}
  {\bibfield  {journal} {\bibinfo  {journal} {arXiv preprint arXiv:1808.04930}\
  } (\bibinfo {year} {2018})}\BibitemShut {NoStop}%
\bibitem [{\citenamefont {Tompson}\ \emph {et~al.}(2016)\citenamefont
  {Tompson}, \citenamefont {Schlachter}, \citenamefont {Sprechmann},\ and\
  \citenamefont {Perlin}}]{tompson2016accelerating}%
  \BibitemOpen
  \bibfield  {author} {\bibinfo {author} {\bibfnamefont {J.}~\bibnamefont
  {Tompson}}, \bibinfo {author} {\bibfnamefont {K.}~\bibnamefont {Schlachter}},
  \bibinfo {author} {\bibfnamefont {P.}~\bibnamefont {Sprechmann}}, \ and\
  \bibinfo {author} {\bibfnamefont {K.}~\bibnamefont {Perlin}},\ }\href@noop {}
  {\bibfield  {journal} {\bibinfo  {journal} {arXiv preprint arXiv:1607.03597}\
  } (\bibinfo {year} {2016})}\BibitemShut {NoStop}%
\bibitem [{\citenamefont {Xie}\ \emph {et~al.}(2018)\citenamefont {Xie},
  \citenamefont {Franz}, \citenamefont {Chu},\ and\ \citenamefont
  {Thuerey}}]{xie2018tempogan}%
  \BibitemOpen
  \bibfield  {author} {\bibinfo {author} {\bibfnamefont {Y.}~\bibnamefont
  {Xie}}, \bibinfo {author} {\bibfnamefont {E.}~\bibnamefont {Franz}}, \bibinfo
  {author} {\bibfnamefont {M.}~\bibnamefont {Chu}}, \ and\ \bibinfo {author}
  {\bibfnamefont {N.}~\bibnamefont {Thuerey}},\ }\href@noop {} {\bibfield
  {journal} {\bibinfo  {journal} {arXiv preprint arXiv:1801.09710}\ } (\bibinfo
  {year} {2018})}\BibitemShut {NoStop}%
\bibitem [{\citenamefont {de~Bezenac}\ \emph {et~al.}(2017)\citenamefont
  {de~Bezenac}, \citenamefont {Pajot},\ and\ \citenamefont
  {Gallinari}}]{de2017deep}%
  \BibitemOpen
  \bibfield  {author} {\bibinfo {author} {\bibfnamefont {E.}~\bibnamefont
  {de~Bezenac}}, \bibinfo {author} {\bibfnamefont {A.}~\bibnamefont {Pajot}}, \
  and\ \bibinfo {author} {\bibfnamefont {P.}~\bibnamefont {Gallinari}},\
  }\href@noop {} {\bibfield  {journal} {\bibinfo  {journal} {arXiv preprint
  arXiv:1711.07970}\ } (\bibinfo {year} {2017})}\BibitemShut {NoStop}%
\bibitem [{\citenamefont {Rasp}\ \emph {et~al.}(2018)\citenamefont {Rasp},
  \citenamefont {Pritchard},\ and\ \citenamefont {Gentine}}]{rasp2018deep}%
  \BibitemOpen
  \bibfield  {author} {\bibinfo {author} {\bibfnamefont {S.}~\bibnamefont
  {Rasp}}, \bibinfo {author} {\bibfnamefont {M.~S.}\ \bibnamefont {Pritchard}},
  \ and\ \bibinfo {author} {\bibfnamefont {P.}~\bibnamefont {Gentine}},\
  }\href@noop {} {\bibfield  {journal} {\bibinfo  {journal} {arXiv preprint
  arXiv:1806.04731}\ } (\bibinfo {year} {2018})}\BibitemShut {NoStop}%
\bibitem [{\citenamefont {J{\'o}nsson}\ \emph {et~al.}(2018)\citenamefont
  {J{\'o}nsson}, \citenamefont {Bauer},\ and\ \citenamefont
  {Carleo}}]{jonsson2018neural}%
  \BibitemOpen
  \bibfield  {author} {\bibinfo {author} {\bibfnamefont {B.}~\bibnamefont
  {J{\'o}nsson}}, \bibinfo {author} {\bibfnamefont {B.}~\bibnamefont {Bauer}},
  \ and\ \bibinfo {author} {\bibfnamefont {G.}~\bibnamefont {Carleo}},\
  }\href@noop {} {\bibfield  {journal} {\bibinfo  {journal} {arXiv preprint
  arXiv:1808.05232}\ } (\bibinfo {year} {2018})}\BibitemShut {NoStop}%
\bibitem [{\citenamefont {Sorella}(2001)}]{sorella2001generalized}%
  \BibitemOpen
  \bibfield  {author} {\bibinfo {author} {\bibfnamefont {S.}~\bibnamefont
  {Sorella}},\ }\href@noop {} {\bibfield  {journal} {\bibinfo  {journal}
  {Physical Review B}\ }\textbf {\bibinfo {volume} {64}},\ \bibinfo {pages}
  {024512} (\bibinfo {year} {2001})}\BibitemShut {NoStop}%
\bibitem [{\citenamefont {Vidal}(2004)}]{vidal2004efficient}%
  \BibitemOpen
  \bibfield  {author} {\bibinfo {author} {\bibfnamefont {G.}~\bibnamefont
  {Vidal}},\ }\href@noop {} {\bibfield  {journal} {\bibinfo  {journal}
  {Physical review letters}\ }\textbf {\bibinfo {volume} {93}},\ \bibinfo
  {pages} {040502} (\bibinfo {year} {2004})}\BibitemShut {NoStop}%
\bibitem [{\citenamefont {Nesterov}(1983)}]{nesterov1983method}%
  \BibitemOpen
  \bibfield  {author} {\bibinfo {author} {\bibfnamefont {Y.~E.}\ \bibnamefont
  {Nesterov}},\ }in\ \href@noop {} {\emph {\bibinfo {booktitle} {Dokl. Akad.
  Nauk SSSR}}},\ Vol.\ \bibinfo {volume} {269}\ (\bibinfo {year} {1983})\ pp.\
  \bibinfo {pages} {543--547}\BibitemShut {NoStop}%
\bibitem [{\citenamefont {Sutskever}\ \emph {et~al.}(2013)\citenamefont
  {Sutskever}, \citenamefont {Martens}, \citenamefont {Dahl},\ and\
  \citenamefont {Hinton}}]{sutskever2013importance}%
  \BibitemOpen
  \bibfield  {author} {\bibinfo {author} {\bibfnamefont {I.}~\bibnamefont
  {Sutskever}}, \bibinfo {author} {\bibfnamefont {J.}~\bibnamefont {Martens}},
  \bibinfo {author} {\bibfnamefont {G.}~\bibnamefont {Dahl}}, \ and\ \bibinfo
  {author} {\bibfnamefont {G.}~\bibnamefont {Hinton}},\ }in\ \href@noop {}
  {\emph {\bibinfo {booktitle} {International conference on machine
  learning}}}\ (\bibinfo {year} {2013})\ pp.\ \bibinfo {pages}
  {1139--1147}\BibitemShut {NoStop}%
\bibitem [{\citenamefont {Kingma}\ and\ \citenamefont
  {Ba}(2014)}]{kingma2014adam}%
  \BibitemOpen
  \bibfield  {author} {\bibinfo {author} {\bibfnamefont {D.~P.}\ \bibnamefont
  {Kingma}}\ and\ \bibinfo {author} {\bibfnamefont {J.}~\bibnamefont {Ba}},\
  }\href@noop {} {\bibfield  {journal} {\bibinfo  {journal} {arXiv preprint
  arXiv:1412.6980}\ } (\bibinfo {year} {2014})}\BibitemShut {NoStop}%
\bibitem [{\citenamefont {Schwarz}\ \emph {et~al.}(2017)\citenamefont
  {Schwarz}, \citenamefont {Alavi},\ and\ \citenamefont
  {Booth}}]{schwarz2017projector}%
  \BibitemOpen
  \bibfield  {author} {\bibinfo {author} {\bibfnamefont {L.~R.}\ \bibnamefont
  {Schwarz}}, \bibinfo {author} {\bibfnamefont {A.}~\bibnamefont {Alavi}}, \
  and\ \bibinfo {author} {\bibfnamefont {G.~H.}\ \bibnamefont {Booth}},\
  }\href@noop {} {\bibfield  {journal} {\bibinfo  {journal} {Physical review
  letters}\ }\textbf {\bibinfo {volume} {118}},\ \bibinfo {pages} {176403}
  (\bibinfo {year} {2017})}\BibitemShut {NoStop}%
\bibitem [{\citenamefont {Sabzevari}\ and\ \citenamefont
  {Sharma}(2018)}]{sabzevari2018faster}%
  \BibitemOpen
  \bibfield  {author} {\bibinfo {author} {\bibfnamefont {I.}~\bibnamefont
  {Sabzevari}}\ and\ \bibinfo {author} {\bibfnamefont {S.}~\bibnamefont
  {Sharma}},\ }\href@noop {} {\bibfield  {journal} {\bibinfo  {journal} {arXiv
  preprint arXiv:1807.10633}\ } (\bibinfo {year} {2018})}\BibitemShut {NoStop}%
\bibitem [{\citenamefont {Xu}\ \emph {et~al.}(2018)\citenamefont {Xu},
  \citenamefont {He}, \citenamefont {De~Sa}, \citenamefont {Mitliagkas},\ and\
  \citenamefont {Re}}]{xu2018accelerated}%
  \BibitemOpen
  \bibfield  {author} {\bibinfo {author} {\bibfnamefont {P.}~\bibnamefont
  {Xu}}, \bibinfo {author} {\bibfnamefont {B.}~\bibnamefont {He}}, \bibinfo
  {author} {\bibfnamefont {C.}~\bibnamefont {De~Sa}}, \bibinfo {author}
  {\bibfnamefont {I.}~\bibnamefont {Mitliagkas}}, \ and\ \bibinfo {author}
  {\bibfnamefont {C.}~\bibnamefont {Re}},\ }in\ \href@noop {} {\emph {\bibinfo
  {booktitle} {International Conference on Artificial Intelligence and
  Statistics}}}\ (\bibinfo {year} {2018})\ pp.\ \bibinfo {pages}
  {58--67}\BibitemShut {NoStop}%
\bibitem [{\citenamefont {Toulouse}\ and\ \citenamefont
  {Umrigar}(2007)}]{toulouse2007optimization}%
  \BibitemOpen
  \bibfield  {author} {\bibinfo {author} {\bibfnamefont {J.}~\bibnamefont
  {Toulouse}}\ and\ \bibinfo {author} {\bibfnamefont {C.~J.}\ \bibnamefont
  {Umrigar}},\ }\href@noop {} {\bibfield  {journal} {\bibinfo  {journal} {The
  Journal of chemical physics}\ }\textbf {\bibinfo {volume} {126}},\ \bibinfo
  {pages} {084102} (\bibinfo {year} {2007})}\BibitemShut {NoStop}%
\bibitem [{\citenamefont {Umrigar}\ \emph {et~al.}(2007)\citenamefont
  {Umrigar}, \citenamefont {Toulouse}, \citenamefont {Filippi}, \citenamefont
  {Sorella},\ and\ \citenamefont {Hennig}}]{umrigar2007alleviation}%
  \BibitemOpen
  \bibfield  {author} {\bibinfo {author} {\bibfnamefont {C.}~\bibnamefont
  {Umrigar}}, \bibinfo {author} {\bibfnamefont {J.}~\bibnamefont {Toulouse}},
  \bibinfo {author} {\bibfnamefont {C.}~\bibnamefont {Filippi}}, \bibinfo
  {author} {\bibfnamefont {S.}~\bibnamefont {Sorella}}, \ and\ \bibinfo
  {author} {\bibfnamefont {R.~G.}\ \bibnamefont {Hennig}},\ }\href@noop {}
  {\bibfield  {journal} {\bibinfo  {journal} {Physical review letters}\
  }\textbf {\bibinfo {volume} {98}},\ \bibinfo {pages} {110201} (\bibinfo
  {year} {2007})}\BibitemShut {NoStop}%
\bibitem [{\citenamefont {Neuscamman}\ \emph {et~al.}(2012)\citenamefont
  {Neuscamman}, \citenamefont {Umrigar},\ and\ \citenamefont
  {Chan}}]{neuscamman2012optimizing}%
  \BibitemOpen
  \bibfield  {author} {\bibinfo {author} {\bibfnamefont {E.}~\bibnamefont
  {Neuscamman}}, \bibinfo {author} {\bibfnamefont {C.}~\bibnamefont {Umrigar}},
  \ and\ \bibinfo {author} {\bibfnamefont {G.~K.-L.}\ \bibnamefont {Chan}},\
  }\href@noop {} {\bibfield  {journal} {\bibinfo  {journal} {Physical Review
  B}\ }\textbf {\bibinfo {volume} {85}},\ \bibinfo {pages} {045103} (\bibinfo
  {year} {2012})}\BibitemShut {NoStop}%
\bibitem [{\citenamefont {LeCun}\ \emph {et~al.}(1989)\citenamefont {LeCun},
  \citenamefont {Boser}, \citenamefont {Denker}, \citenamefont {Henderson},
  \citenamefont {Howard}, \citenamefont {Hubbard},\ and\ \citenamefont
  {Jackel}}]{lecun1989backpropagation}%
  \BibitemOpen
  \bibfield  {author} {\bibinfo {author} {\bibfnamefont {Y.}~\bibnamefont
  {LeCun}}, \bibinfo {author} {\bibfnamefont {B.}~\bibnamefont {Boser}},
  \bibinfo {author} {\bibfnamefont {J.~S.}\ \bibnamefont {Denker}}, \bibinfo
  {author} {\bibfnamefont {D.}~\bibnamefont {Henderson}}, \bibinfo {author}
  {\bibfnamefont {R.~E.}\ \bibnamefont {Howard}}, \bibinfo {author}
  {\bibfnamefont {W.}~\bibnamefont {Hubbard}}, \ and\ \bibinfo {author}
  {\bibfnamefont {L.~D.}\ \bibnamefont {Jackel}},\ }\href@noop {} {\bibfield
  {journal} {\bibinfo  {journal} {Neural computation}\ }\textbf {\bibinfo
  {volume} {1}},\ \bibinfo {pages} {541} (\bibinfo {year} {1989})}\BibitemShut
  {NoStop}%
\bibitem [{\citenamefont {Cai}\ and\ \citenamefont
  {Liu}(2018)}]{cai2018approximating}%
  \BibitemOpen
  \bibfield  {author} {\bibinfo {author} {\bibfnamefont {Z.}~\bibnamefont
  {Cai}}\ and\ \bibinfo {author} {\bibfnamefont {J.}~\bibnamefont {Liu}},\
  }\href@noop {} {\bibfield  {journal} {\bibinfo  {journal} {Physical Review
  B}\ }\textbf {\bibinfo {volume} {97}},\ \bibinfo {pages} {035116} (\bibinfo
  {year} {2018})}\BibitemShut {NoStop}%
\bibitem [{\citenamefont {Peruzzo}\ \emph {et~al.}(2014)\citenamefont
  {Peruzzo}, \citenamefont {McClean}, \citenamefont {Shadbolt}, \citenamefont
  {Yung}, \citenamefont {Zhou}, \citenamefont {Love}, \citenamefont
  {Aspuru-Guzik},\ and\ \citenamefont {O’brien}}]{peruzzo2014variational}%
  \BibitemOpen
  \bibfield  {author} {\bibinfo {author} {\bibfnamefont {A.}~\bibnamefont
  {Peruzzo}}, \bibinfo {author} {\bibfnamefont {J.}~\bibnamefont {McClean}},
  \bibinfo {author} {\bibfnamefont {P.}~\bibnamefont {Shadbolt}}, \bibinfo
  {author} {\bibfnamefont {M.-H.}\ \bibnamefont {Yung}}, \bibinfo {author}
  {\bibfnamefont {X.-Q.}\ \bibnamefont {Zhou}}, \bibinfo {author}
  {\bibfnamefont {P.~J.}\ \bibnamefont {Love}}, \bibinfo {author}
  {\bibfnamefont {A.}~\bibnamefont {Aspuru-Guzik}}, \ and\ \bibinfo {author}
  {\bibfnamefont {J.~L.}\ \bibnamefont {O’brien}},\ }\href@noop {} {\bibfield
   {journal} {\bibinfo  {journal} {Nature communications}\ }\textbf {\bibinfo
  {volume} {5}},\ \bibinfo {pages} {4213} (\bibinfo {year} {2014})}\BibitemShut
  {NoStop}%
\bibitem [{\citenamefont {Zoph}\ and\ \citenamefont
  {Le}(2016)}]{zoph2016neural}%
  \BibitemOpen
  \bibfield  {author} {\bibinfo {author} {\bibfnamefont {B.}~\bibnamefont
  {Zoph}}\ and\ \bibinfo {author} {\bibfnamefont {Q.~V.}\ \bibnamefont {Le}},\
  }\href@noop {} {\bibfield  {journal} {\bibinfo  {journal} {arXiv preprint
  arXiv:1611.01578}\ } (\bibinfo {year} {2016})}\BibitemShut {NoStop}%
\bibitem [{\citenamefont {Pham}\ \emph {et~al.}(2018)\citenamefont {Pham},
  \citenamefont {Guan}, \citenamefont {Zoph}, \citenamefont {Le},\ and\
  \citenamefont {Dean}}]{pham2018efficient}%
  \BibitemOpen
  \bibfield  {author} {\bibinfo {author} {\bibfnamefont {H.}~\bibnamefont
  {Pham}}, \bibinfo {author} {\bibfnamefont {M.~Y.}\ \bibnamefont {Guan}},
  \bibinfo {author} {\bibfnamefont {B.}~\bibnamefont {Zoph}}, \bibinfo {author}
  {\bibfnamefont {Q.~V.}\ \bibnamefont {Le}}, \ and\ \bibinfo {author}
  {\bibfnamefont {J.}~\bibnamefont {Dean}},\ }\href@noop {} {\bibfield
  {journal} {\bibinfo  {journal} {arXiv preprint arXiv:1802.03268}\ } (\bibinfo
  {year} {2018})}\BibitemShut {NoStop}%
\end{thebibliography}%

\newpage

\beginsupplement
\setcounter{section}{0}
\setcounter{equation}{0}
\setcounter{figure}{0}
\setcounter{table}{0}
\setcounter{page}{1}
\makeatletter
\renewcommand{\theequation}{S\arabic{equation}}
\renewcommand{\thefigure}{S\arabic{figure}}

\renewcommand{\thesection}{S\arabic{section}}   

\part*{Supplementary Material}

\section{Additional architectures}
\label{subsec:AdditionalArchitectures}
\begin{figure}[H]
  \includegraphics[width=7cm]{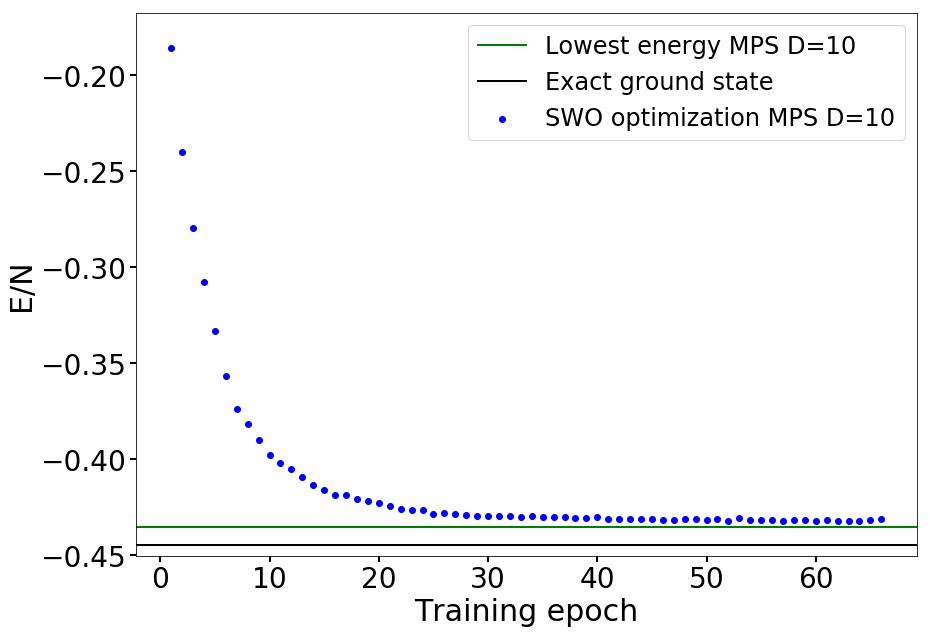}
  \includegraphics[width=7cm]{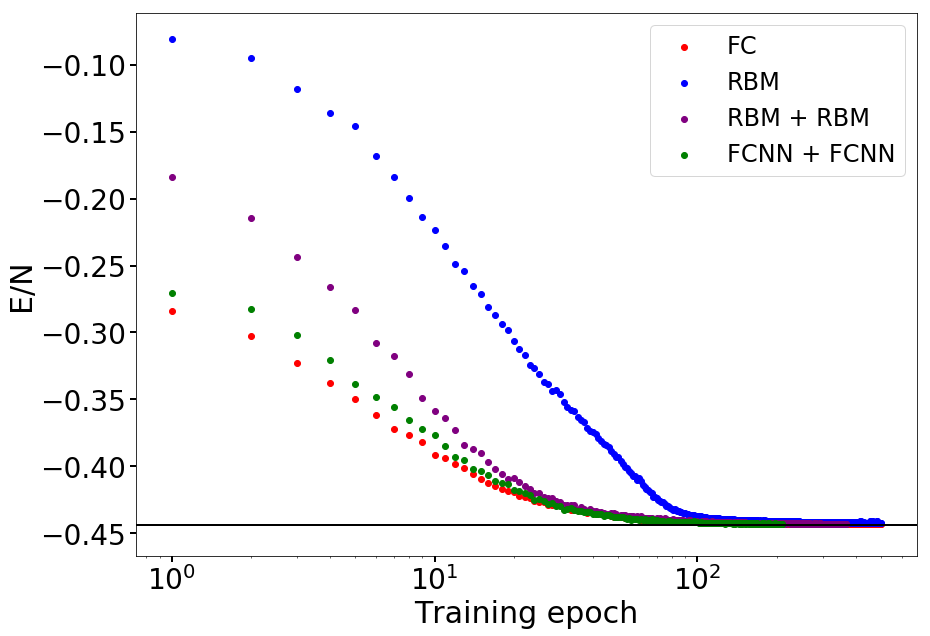}
  \caption{SWO optimization of matrix product state with bond dimension 10 on a 24 site Heisenberg chain (\textit{top}) and sums of wave-functions on a 40 site Heisenberg chain (\textit{bottom}).}
  \label{fig:1d_40_sums}
\end{figure}

In addition to the ansatz mentioned in the main text, we also construct computational graphs for matrix-product states (MPS) and multi-FCNN.   Results of the optimization of matrix-product states is shown in fig~\ref{fig:1d_40_sums}(top).  In practice this is an inefficient way to optimize MPS compared against the density matrix renormalization group (DMRG) which is explicitly designed for this variational ansatz.  On the other hand,  for product ansatz such as Slater-MPS \cite{chou2012matrix} where DMRG is inapplicable, this may be an effective approach. 

In the CGS representation sum architecture, such as multi-FCNN, can be implemented by simple addition of a sum node that combines contributions of individual ansatzs.   Results of optimization for multi-FCNN (shown along with multi-RBM which is also in the main text) are shown in figs.~\ref{fig:1d_40_sums}(bottom).

\section{Optimization regimes}
\label{sec:SuppOptimization}

\begin{figure}[H]
    \includegraphics[width=0.5\textwidth]{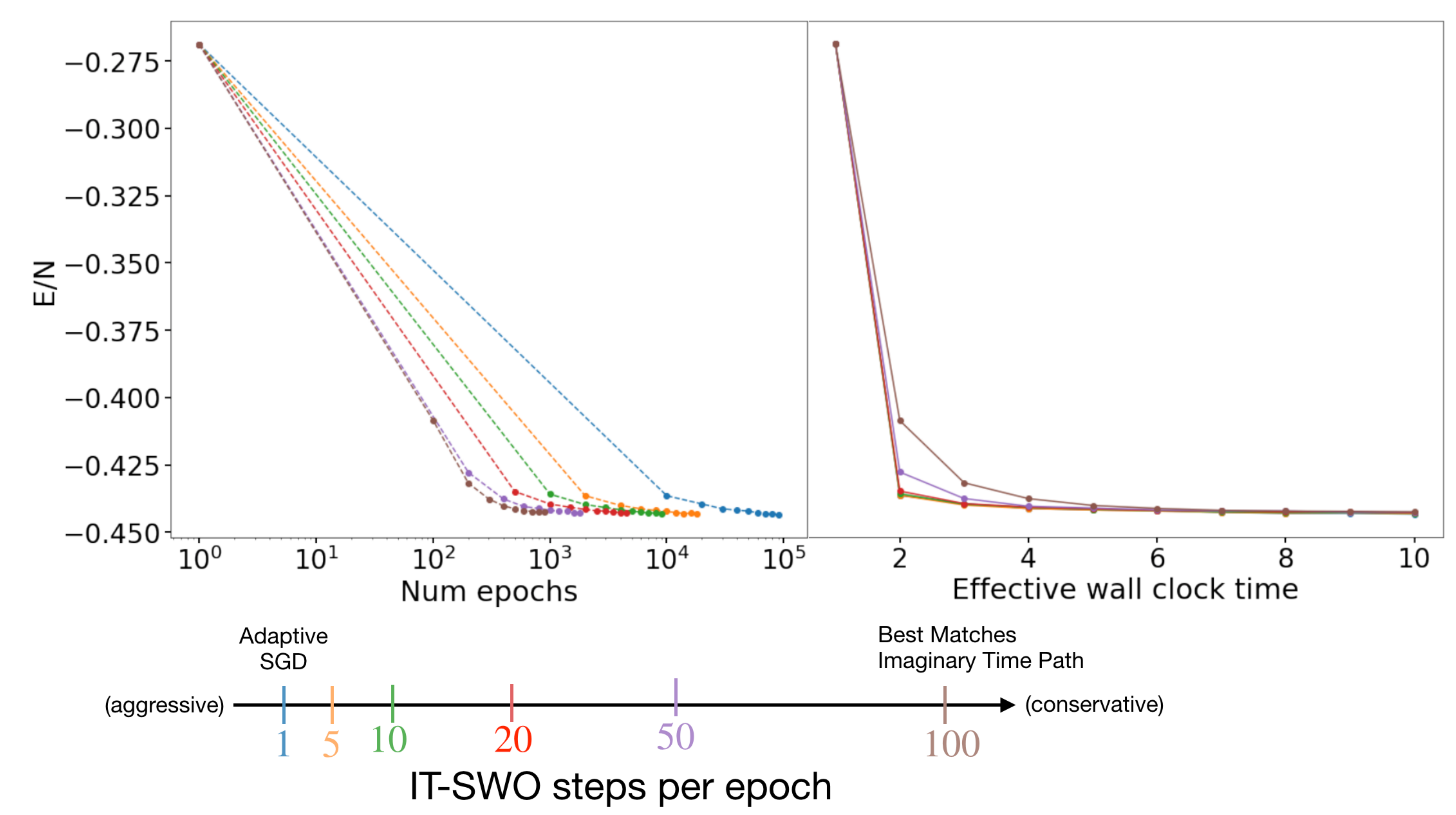}
    \label{fig:steps_per_epoch}
    \caption{Energy per site of 24 site 1D Heisenberg model shown both as a function of epochs and effective wall-clock time for different number of steps per epoch.  Here we minimize the log-overlap\cite{jonsson2018neural}.  Larger steps per epoch decrease total number of epochs but can increase wall-clock time as each epoch is slower. The optimal trade-off is likely problem specific and remains on open question.}
    \label{fig:num_batches_comparison}
\end{figure}

Here we show results of using IT-SWO as an optimization method in different regimes. In fig.~\ref{fig:num_batches_comparison}, we show energy traces as a function of number of epochs and wall clock times.

\begin{figure}[H]
    \includegraphics[width=0.5\textwidth]{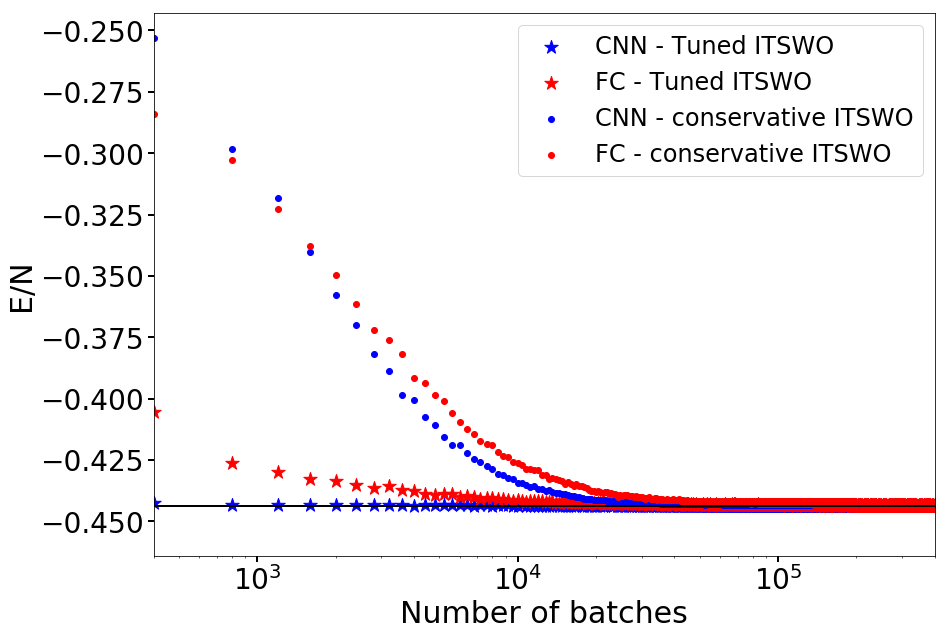}
    \label{fig:tuned_optimization}
    \caption{Optimization of CNN and FC architectures with tuned hyperparameters on 40 site 1D Heisenberg model using LogOverlap ITSWO and adaptive gradient descent. In this setting energy converges to -0.4436 within a 1000 epochs, which takes only several hours on a single GPU enabled machine.}
\end{figure}

\begin{figure}
    \includegraphics[width=7cm]{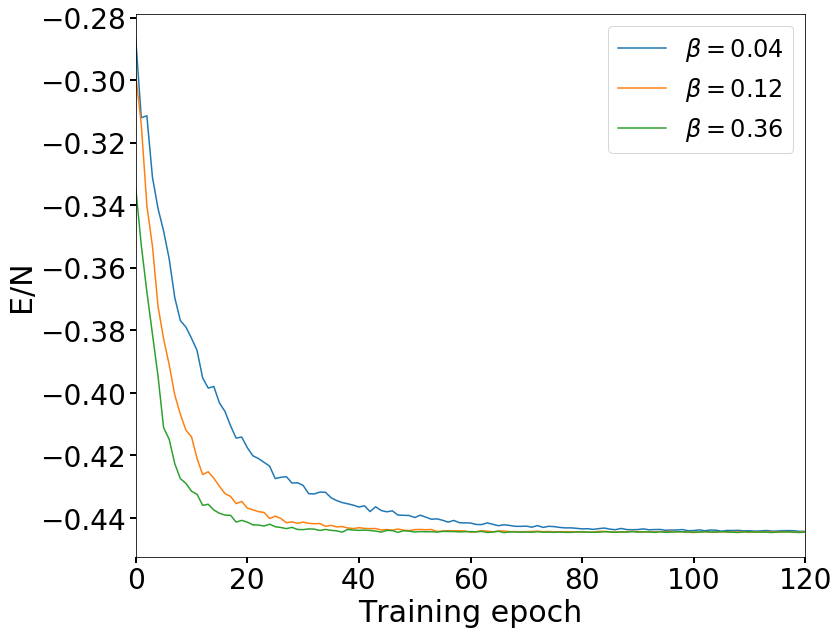}
    \caption{Energy profiles as a function of number of epochs for varying values of $\beta$. In all simulations the number of batches is chosen such as to converge to the next target state for all values of $\beta$. Optimization is performed on 24 sites 1D Heisenberg model with periodic boundary conditions.}
    \label{fig:varying_beta}
\end{figure}

\section{Fidelity maximization}
\label{sec:SuppFidelity}

SWO optimization is based on fidelity maximization. In the main text we used minimization of $\mathcal{L}_{2}(\epsilon_{rr})$ as an objective function. Here we show that fidelity error and $|\epsilon_{rr}|^{2}$ are the same to first order:

\begin{align}
    f_{err} & = 1 - \frac{\braket{\psi_{T}}{\psi_{W}}} {\sqrt{|\psi_{T}|^{2}|\psi_{W}|^{2}}} \\
    		& = 1 - \frac{\braket{\psi_{T}}{\psi_{T} + \epsilon_{rr}}}{\sqrt{|\psi_{T}|^{2} |\psi_{T} + \epsilon_{rr}|^{2}}}\\
    		& = \frac{\sqrt{|\psi_{T}|^{2}|\psi_{T} + \epsilon_{rr}|^{2}} - |\psi_{T}|^{2} - \braket{\psi_{T}}{\epsilon_{rr}}}{\sqrt{|\psi_{T}|^{2} |\psi_{T} + \epsilon_{rr}|^{2}}} \\
            &= \frac{|\psi_{T}|^{2} \sqrt{\frac{|\psi_{T} + \epsilon_{rr}|^{2}}{ |\psi_{T}|^{2}}}  - |\psi_{T}|^{2} - \braket{\psi_{T}}{\epsilon_{rr}}}{\sqrt{|\psi_{T}|^{2} |\psi_{T} + \epsilon_{rr}|^{2}}} \\
            & \approx \frac{|\epsilon_{rr}|^{2} }{2|\psi_{T}|^{2}} \sim |\epsilon_{rr}|^{2}
\end{align}

In addition we can show that minimization of $\mathcal{L}_{2}(\psi-\phi)$ provides a good proxy for fidelity maximization even for arbitrary wave-functions. We do that by showing that it establishes an upper bound on the fidelity error and hence its minimization acts as a contractive map. Assuming that $|\phi| > |\psi|$ without loss of generality
    
\begin{align}
f_{err} & = 1 - \frac{\braket{\psi}{\phi}}{\sqrt{|\psi|^{2}|\phi|^{2}}} \\
		& = \frac{\sqrt{|\psi|^{2}|\phi|^{2}} - \sum\limits_{i}\psi_{i} \phi_{i}}{\sqrt{|\psi|^{2}|\phi|^{2}}} \\
        & \leq \frac{|\phi|^{2} - \sum\limits_{i}\psi_{i} \phi_{i}}{\sqrt{|\psi|^{2}|\phi|^{2}}} \\
        & = \frac{\sum\limits_{i}\phi_{i} \phi_{i} - \sum\limits_{i}\psi_{i} \phi_{i}}{\sqrt{|\psi|^{2}|\phi|^{2}}} \\
        & = \frac{\sum\limits_{i}\phi_{i} (\phi_{i} - \psi_{i})}{\sqrt{|\psi|^{2}|\phi|^{2}}} \label{schwartz}\\
        & \leq \frac{\sqrt{\sum\limits_{i}\phi_{i}^{2}} \sqrt{(\phi_{i} - \psi_{i})^{2}}}{\sqrt{|\psi|^{2}|\phi|^{2}}} \\
        & = \frac{\sqrt{\mathcal{L_{2}}(\phi_{i} - \psi_{i})}}{\sqrt{|\psi|^{2}}}
\end{align}
Given 
\begin{equation}
\mathcal{L_{2}}(\tilde{\psi} - \tilde{\phi}) \leq \epsilon 
\end{equation}
by working with unnormalized wavefunctions with a scaling factor that insures $|\psi| > 1$ we conclude
\begin{equation}
f_{err} \leq \sqrt{\epsilon}
\end{equation}
Here on line \ref{schwartz} we used Cauchy-Schwartz inequality.

An alternative approach to fidelity maximization \cite{jonsson2018neural} is to minimize the negative $\log$ overlap. In the limit where $\ket{\psi_T}$ is close to $\ket{\psi_W}$ this approach is equivalent to minimizing $\mathcal{L}_{2}(\epsilon_{rr})$. While working with $\mathcal{L}_{2}(\epsilon_{rr})$ provides an estimate of the optimization process and can be directly used to adaptively choose the number of steps per epoch, minimization of $-\log(f)$ avoids the complications arising from normalization.
At large $\beta$ and single step per epoch the log-overlap IT-SWO matches adaptive SGD exactly and does not have the $\beta^2$ difference from the standard approach to normalization.

\section{SWO Sampling}

In this section we discuss alternative sampling methods for SWO. 
  
In the main text we described a procedure where weighted $\mathcal{L}_{2}(\psi_{W} - \psi_{T})$ is minimized on configurations sampled from $|\psi_{W}|^{2}$. This choice allows us to easily compute energy expectation values and draw gradients from largest wave-function amplitudes. This approach is very similar to the traditional VMC optimization.  While a valuable approach, it is important to accurately clean up small-amplitude configurations toward the end of the optimization and this is difficult when primarily sampling large-amplitude states.   One approach to tackle this problem is to sample configurations that are likely to provide largest gradients, i.e. sample from $|\psi_{W} - \psi_{T}|^{2}$. While this approach theoretically provides clearer signal, in practice our experiments using it did not successfully improve the result.  We believe this is because, while being a better distribution to sample, standard Monte Carlo moves are not effective at sampling from it.

\section{TensorFlow implementation}
\label{sec:TensorFlow}
In this work we have designed the entire optimization process to fit into the framework of computational graphs. The code developed as a part of this work, CGS-VMC, is provided as an open source project on Github (https://github.com/ClarkResearchGroup/cgs-vmc). Here we note several aspects on the implementation.

Since the optimization phase of SWO is based on supervised training it naturally fits into the machine learning framework. Other operations, such as Monte Carlo sampling and system specific Markov chain updates were implemented using existing primitives. For example, the traditional exchange update of two spins in opposite configurations is implemented through the process of scaling of all spins by random amplitudes and swapping the largest and smallest entries. Complexity of such operations can be further reduced by integrating custom implementations into TensorFlow. Despite suboptimal implementation of particular operations we find that most of the computational resource are spent on the wave-function evaluation.

Besides the flexibility of designing novel variational architectures, building wave-function optimization code around TensorFlow enables us to trivially utilize GPU accelerators for optimization. This is a significant advantage: for comparison, a single optimization epoch of a CNN model on 64 site square lattice takes on the order of 30 minutes on CPU and only 1-2 minutes when executed on GPU.

It's worth mentioning that some variational wave-functions allow efficient update formulas. In most cases this is based on reuse of unaffected parts of computation (e.g. minors in the determinant evaluation). Incorporating such optimizations would further speed up our framework, but was not prioritized as most of our ansatzs involve multiple nonlinear transformations making such updates inapplicable. 

During traditional VMC optimization samples on which gradients are evaluated are drawn from an equilibrated Markov chain. To keep track of this stateful component, we use non-trainable variables, which we iteratively update based on Metropolis-Hastings rule throughout the optimization. This is reflected in the execution schedule, where every training iteration is separated by Monte Carlo sweep.

\section{Evaluation of the normalization constant}
As mentioned in the main text, in the imaginary time evolution instantiation of SWO we introduce a normalization constant $N = (1 - 2 \beta \langle E \rangle + \beta^{2} \langle E^{2} \rangle)^{2}$ to account for non-unitarity of $1 - \beta \hat{H}$. We can estimate this value during training by computing expectation values of $\langle E \rangle$ and $\langle E^{2} \rangle$. While this could be done independently, this quantity can be readily estimated from $\braket{c}{\psi_{W}}$ and $\braket{c}{\psi_{T}}$

\begin{align}
    &\langle E \rangle = \sum_{c ~|\psi_{W}|^{2}} \frac{\braket{c}{\hat{H}\psi_{W}}}{\braket{c}{\psi_{W}}} \\
    &\langle E^{2} \rangle =  \sum_{c} \frac{ \braket{c}{\hat{H}\psi_{W}} \braket{c}{\hat{H}\psi_{W}}}{\braket{c}{\psi_{W}} \braket{c}{\psi_{W}}} \\
\end{align}

The quantity $\braket{c}{\hat{H}\psi_{W}}$ is a part of the $\psi_{T}$. To reduce the variance of $\langle E \rangle$ and $\langle E^{2} \rangle$ we maintain an exponentially moving average of these quantities. For most simulations we used decay rate of $0.999$. While this approach produces normalization that is lagging behind by one epoch, in our numerical experiments we have not observed any significant adverse effects.

\end{document}